Toward a Universal Cortical Algorithm:

Examining Hierarchical Temporal Memory in Light of Frontal Cortical Function

Michael R. Ferrier

Department of Cognitive Linguistic and Psychological Sciences

Brown University

The human brain carries out a broad array of functions, and there is ample evidence that it is the cerebral cortex, taking up three quarters of the brain's volume, that plays a central role in perception, cognition, and behavior. However, notwithstanding the variety and complexity of the tasks performed by the cerebral cortex, its structure is fairly regular (Hubel & Wiesel, 1968; Mountcastle, 1978; Rockel et al, 1980; Mountcastle, 1997; Hirsch & Martinez; 2006; however, see DeFelipe et al, 2002; Herculano-Housel et al, 2008). All areas of cortex have a laminar structure that is typically divided into six layers, each with stereotypical patterns of connectivity and cell type concentrations. Crossing perpendicular to the layer structure are vertically aligned patterns of connectivity by which groups of ~100 cells are densely interconnected. These columns (also known as minicolumns) are often theorized to make up cortical "micro-circuits", individual functional units. Cells within a single minicolumn typically have similar receptive fields and response properties, leading to the hypothesis that each minicolumn acts as an individual pattern recognizer or feature detector. Connectivity and receptive field characteristics also indicate a larger columnar organization, often called the hypercolumn, which is made up of ~100 minicolumns. Inhibitory connections between the minicolumns within a single hypercolumn supress activity in all but those minicolumns that best match their current pattern of input, which facilitates the self-organization of the minicolumns' receptive fields (Lucke and Bouecke, 2005) and produces a sparse re-coding of the input pattern. Different areas of cortex vary in their patterns of intra-cortical connectivity, in their inputs from thalamic and other subcortical nuclei, in the density of particular cell types, in the thickness of cortical layers, etc., but the similarities in architecture between cortical areas and between species is so great as to suggest a common underlying structural and functional template that is repeated many times, with some modification, throughout the cortex.

In addition to structural homogeneity, there are other lines of evidence indicating a common cortical algorithm that is utilized in a domain general manner. Research on statistical learning (Saffran et al., 1996a, 1996b; Saffran et al., 1999; Kirkham et al., 2002; Fiser & Aslin 2002, 2005; Saffran & Wilson, 2003, Ferrier, 2006; Graf Estes et al., 2007) has shown that given structured stimuli composed of visual images, speech sounds, or tones, both infants and adults will learn the underlying structure in a syntactic (ie., compositional and hierarchical) manner. Results across the various domains show very similar properties; input is broken up into statistically coherent chunks at the "fault lines" of lower predictive probability, and these learned chunks are then made available as components for higher level associative learning.

There is also evidence of a common cortical algorithm from surgical manipulation. Functional interchangeability has been demonstrated by an experiment in which visual input was surgically rerouted to auditory cortex in neonatal ferrets, and the mature animals were able to respond to visual stimuli, with retinotopic maps and typical visual receptive fields having developed within their auditory cortex (von Melchner et al., 2000).

The plasticity which allows one cortical area to take over the function of a second area has been studied extensively (e.g., Feldman, 2009). In a recent example, it was shown that the visual cortex of congenitally blind individuals will often come to respond to language (Bedny et al., 2011). BOLD imaging and functional connectivity indicate that these visual areas become involved in language processing, of the same type normally seen only in specific areas of the left frontal and temporal lobes. Vision and language have often been considered two of the domains most likely to be processed by specialized cortical circuits that have evolved for their particular functions (e.g., Pinker, 1994; Callaway, 1998), and yet functional interchangeability is now being seen even between cortical areas normally dedicated to these domains.

**Constraints on a Universal Cortical Algorithm**

The prospect that there may be a single underlying process that the cortex applies many millions of times, in parallel, and in a domain general manner, is extremely attractive. It presents the possibility of understanding in one stroke how the cortex plays its central roles in perception, cognition and behavior. However, this also places an enormous explanatory burden upon any candidate theory for such a cortical algorithm. In recent years there have been an increasing number of attempts to synthesize the growing body of knowledge of the details of cortical function into a coherent theory (e.g., Grossberg 2007, Rodriguez et al., 2005; Hecht-Nielsen, 2007; Bastos et al., 2012), but no such theory has yet materialized that is clearly able to explain the full diversity of cortical functions. Nonetheless, many constraints and hints have emerged as to the shape this common algorithm may take, both from the computational level (Marr, 1982), in the form of mathematical models of learning and information processing, and from the implementation level, in the form of the advancing knowledge of cortical neurophysiology, neuroanatomy and connectivity. In this section, I will review several diverse but complementary areas of research that, taken together, begin to paint a picture of how a universal cortical algorithm may work.

**Sparse Distributed Representation**

One of the most fundamental aspects of cortical function is the way in which knowledge about the world is represented by the activity of cells within a region of cortex. The spectrum of possibilities begins with the localist representation, in which the activity of a single cell corresponds to a given percept or concept. At the other end of the spectrum is the dense distributed representation, in which the activity of a large fraction (up to ~50%) of cells within an area acts as a representation. Between these two extremes is the sparse distributed representation, in which the activity of a small fraction of the cell population within an area acts as a representation.

Distributed representation offers many advantages over localist representation (Hinton, 1986). Because distributed representation is combinatorial, storage capacity increases exponentially as the number of units grows; with localist representation, storage capacity increases only linearly with the number of units. Distributed representation facilitates generalization among representations based on the degree of overlap between their respective patterns of activity. The activity of the individual units that make up a distributed representation may have meaning on their own, which allows for semantically relevant overlap between various representations. For example, a representation's pattern of activity may be made up of subpatterns that represent classes in a type hierarchy, or sub-parts in a compositional relationship. With distributed representation, new representations can differentiate themselves progressively by gradual weight modifications, whereas localist representations are formed abruptly and discretely. Finally, by virtue of their redundancy, distributed representations are more fault tolerant than localist representations.

Among distributed representations, a sparse distribution of activity confers most of the same advantages as a dense distribution, while avoiding several of its drawbacks (Foldiak, 2002; Willshaw and Dayan, 1990). While the representational capacity of a sparse representation is very high, the representational capacity of a dense representation is much higher still — in fact unnecessarily high, resulting in a great degree of redundancy between representations. The high information content of dense distributed representations make associative mappings between representations much more complex. The mappings would generally not be linearly separable, and so would require multilayer networks and learning algorithms that strain biological plausibility. Linear separability is much more readily achieved with sparse distributed representations, which allows many more associative pairs to be learned by a single layer of connections modified by a biologically plausible local Hebbian learning function (Olshausen & Field, 2004). Both types of distributed representation support generalization between overlapping patterns, but because the activity of a single unit in a sparse representation has lower probability and can therefore be more selective, there is greater opportunity (depending on the learning algorithm) for a sparse representation to mix several semantically meaningful subpatterns in a combinatorial manner.

Sparse distributed representations, therefore, have a number of computational advantages over the alternatives. And in fact, studies have begun to reveal cortical cell response properties consistent with sparse distributed representations (Olshausen & Field, 2004). Sparse patterns of activity corresponding with sensory input have been observed in rat auditory and somatosensory cortex and in the insect olfactory system. Sparse activity has been recorded in motor cortex accompanying the initiation of movement. Sparse activity has also been observed in primate V1 when the subject is exposed to natural visual scenes. Interestingly, when provided with stimulation only within their own receptive fields, these same V1 cells become more densely active, indicating that context affects the degree of sparseness of activity. This point will be relevant to the discussions in later sections of predictive coding and hierarchical temporal memory.

The low probability of activation and narrow selectivity of the preferred stimulus of individual cells in higher visual areas support the hypothesis that neural representations there are sparse, and that elements of those representations are meaningful in and of themselves, representing e.g. complex shapes, object components, and faces (Gross et al., 1972; Perrett et al., 1982; Tanaka, 1996; Foldiak, 2002). Zhang et al. (2011) recorded responses from ~200 monkey IT cells and used a classifier to identify the sparse activity patterns corresponding to each of several visual objects shown in isolation. These patterns carried information specific to both the identity and position of the objects. When several of these objects were then displayed simultaneously and peripherally, the resulting pattern of activity contained information about the identity and position of each of the objects, though in a reduced form compared with when the objects were shown in isolation. When attention was then covertly directed to only one of the several objects being displayed, cell activity representing information about that object increased, at the expense of activity corresponding to the unattended objects. These results indicate sparse representation of object identity and position in IT, with representations of individual sub-parts of a scene being combined compositionally, and with the degree of representation of these various sub-parts being modulated by attention.

Computational models have also supported the hypothesis that sparse distributed representations are used by the cortex. Field (1987) and Olshausen and Field (1996) found that when the receptive fields of an entire group of units is optimized to produce sparse representations when trained on natural visual scenes, the units develop response properties similar to those of V1 simple cells. Sparse coding also meets the brain's energy efficiency requirements; Lennie (2003) found that given its energy usage constraints, less than 2% of the brain's neurons could afford to be significantly active at any given time.

**Bayesian Inference**

Visual perception has traditionally been considered mainly in terms of feedforward processing, with increasingly complex and abstract representations at each level building upon the activity of the previous level, while top-down feedback is primarily relegated to the modulation of attention (Marr, 1982; Felleman & Van Essen, 1991; Desimone & Duncan, 1995). Alternately, a number of theories have proposed that feedback may instead provide contextual priors to influence inference at lower levels (McClelland & Rumelhart, 1980; Carpenter & Grossberg, 1987; Mumford, 1992; Rao & Ballard, 1997). If this is the case, then attention, seen as biased competition within a cortical region, may be just one aspect of a process of biased inference that is mediated by top-down feedback. Building on pattern theory (Grenander, 1993), Lee and Mumford (2002) proposed that a cortical hierarchy performs Bayesian belief propagation. In this view, a cortical area treats its bottom-up input (from sensory thalamus or from hierarchically lower cortical areas) as evidence, which it combines with top-down inputs from higher cortical areas that are treated as Bayesian contextual priors, in order to determine its most probable hypotheses and to activate their representations. Over time, the region moves toward an equilibrium in which the optimum set of representations is activated, in order to maximize the probability of the active representations given both the bottom up and top down data. This view has received significant experimental support. In cases where direct input differs from perception, as with binocular rivalry or illusory contours, or where input is ambiguous and may be perceived in more than one way, short latency (100ms-200ms) responses in V1 correspond well with bottom up thalamic input, while longer latency responses instead correspond partially with what is perceived (Lee & Mumford, 2002). The correspondence of relevant activity with what is perceived, rather than with direct input, increases at higher levels of the hierarchy, starting at 10% in V1 and 20% in V2, through to nearly 100% in IT (Logothetis, 1998). The Bayesian belief propagation hypothesis makes sense of these result and others, while providing a biologically plausible framework through which the cortex may implement hierarchical Bayesian inference.

**Deep Learning**

Deep learning architectures (Hinton, 2006; Bengio, 2009) exploit the assumption that the generating causes underlying observations about the world are organized compositionally or categorically into multiple hierarchical levels, and that higher level representations can build upon combinations and transformations of lower level representations. These architectures generally use a combination of supervised and unsupervised learning to extract statistical regularities at each level, passing the transformed input representation up to the next level where further regularities may be extracted. Each layer of the network acts as a filter, capturing a subset of regularities from the input signal, and so reducing the dimensionality of the data. For example, when applied to visual object recognition problems, a deep learning architecture may learn representations at the lowest level that correspond to lines and edges; at the next higher level it may learn representations corresponding to corners and intersections of lines; at the next level it may represent parts of objects, and so on. Deep learning networks have recently produced exceptional results with several classes of pattern recognition tasks, in some cases achieving human-level performance for the first time (e.g., Ciresan et al., 2012a/2012b/2012c). They are also modeled on cortical processing, which uses a hierarchically organized series of regions to extract increasingly compositionally complex and categorically abstract representations from perceptual input.

**Multi-Stage Hubel Wiesel Architectures**

Hierarchical models of cortical function have their roots in the pioneering work of Hubel and Wiesel (1962, 1968), who studied cell response properties in primary visual cortex (V1). They identified two basic patterns of response properties in V1 cells; simple cells respond mostly to lines of a particular orientation, spatial frequency, and position, while complex cells introduce some degree of invariance to position. This pattern was found to be repeated in later regions in the visual hierarchy, with simple cells that would respond to a spatial pattern but with increasing complexity at each level, and complex cells that at each level would introduce a greater degree of invariance to position and scale. A series of models of the ventral visual stream (involved in object recognition) have been developed based on these findings, for which Ranzato et al. (2007) introduced the term multi-stage Hubel-Wiesel architectures (MHWA). These include the Neocognitron (Fukushima, 1988), convolutional networks (Chellapilla et al., 2006; Ciresan et al., 2011), and HMAX (Riesenhuber & Poggio, 1999; Serre et al., 2005/2007a/2007b). MHWA models are composed of alternating layers of conjunctive and disjunctive units. The conjunctive units perform template matching, responding to a particular combination of inputs from the previous layer, such as a contrasting edge at a particular orientation and position (often described using a Gabor filter function). The disjunctive units are wired to respond when any of several related conjunctive units from a local area in the previous layer are active, for example when any unit is active that represents a given line orientation and spatial frequency, but at any position within a local range. Each disjunctive unit pools over a particular set of conjunctive input units. In this way the disjunctive units introduce a measure of position and scale invariance. The disjunctive unit layer then feeds into a second conjunctive unit layer, which learns to respond to specific combinations of those partially invariant representations, and so on. The result is that, over the course of several levels, representations are learned that are selective to individual whole objects but are also invariant to changes in position and scale.

Like other MHWA models, HMAX only attempts to model the feedforward aspect of visual object recognition, representing the first ~100 ms of processing after stimulus onset. Its biologically motivated combination of a deep architecture with the alternating conjunctive and disjunctive layers that introduce position and scale invariance have allowed it to rival the performance of both humans (Serre et al., 2007b) and cutting edge computer vision techniques (Serre et al., 2007a) in fast object recognition tasks. However while HMAX accurately models, and in some cases has predicted (Serre et al., 2005), response properties of several types of cell in visual areas of the cortex (as well as imaging and psychophysical results), it does not address how all of those response properties may develop in the first place. Specifically, HMAX's disjunctive units are hard-wired to the particular set of conjunctive units that they pool over, in order to respond invariantly to the activity of any of those conjunctive units and so introduce spatial and scale invariance; these connections are set parametrically, rather than through a learning process. While some cortical connection strengths are very likely programmed genetically (Markram & Perin, 2011), if cortical complex cells correspond in their properties to HMAX's disjunctive units it may be impractical for their input connections to be formed other than by a learning process, since the response properties of the simple cells that provide their inputs would themselves be learned. The HMAX model remains agnostic to the nature of this learning process.

When considering HMAX as a candidate for a universal model of cortical function, it has several additional shortcomings. The response properties of its pooling disjunctive cells are based closely on the complex cells in visual regions of cortex, and as such are specialized for producing position and scale invariance. It is unclear what role the disjunctive layers may play in non-visual cortical functions. In addition, there are other cortical functions such as the representation of temporal sequences, crucial for auditory and motor processing, that are not modeled at all within the HMAX framework. Even within the domain of visual object recognition, HMAX relies on a separate, supervised mechanism to learn other types of invariance, such as pose, rotation and lighting invariance. HMAX is very successful as a model of certain aspects of the ventral visual stream, but it does not represent a universal model of cortical function.

**Temporal Slowness**

Connectionist models are typically trained either in an unsupervised manner, often using some form of Hebbian learning to produce associations between inputs that co-occur, or else in a supervised manner, using an error driven gradient descent learning method such as backpropagation (Rumelhart et al, 1986) to produce associations between arbitrary pairs of inputs (O'Reilly & Munakata, 2000). One reason visual invariance problems are difficult is because they are not easily solved by either of these methods. Two input patterns representing the same object with some variation in position (or scale, rotation, etc.) may be separated by a very large Euclidean distance, while two patterns representing different objects may be much closer. Hebbian learning is very sensitive to shared structure, and so would be unable to make the correct classifications in such a case. Error driven learning has been proven more capable of making the correct discriminations, but is not very good at generalizing these discriminations beyond its training set. An error driven convolutional network requires a large number of training examples per class (Ranzato et al., 2007), whereas the primate visual system can often discriminate similar objects after just a few training examples. Error driven learning, being supervised, is also arguably not biologically plausible as a form of basic perceptual learning, because there are no category labels in the real world to act as supervisor.

In recent years several new approaches have made headway toward solving this problem, using the principle of temporal slowness (Berkes, 2005; Földiák, 1991; Mitchison, 1991). Temporal slowness is based on the idea that sensory data changes on a much more rapid timescale than do the relevant properties of the world. While the identities of nearby objects and their configuration in the environment tend to change only gradually, the visual input reflecting that environment can change markedly from one moment to the next, due for example to the movement of objects or a change in view angle. Ideally the brain's internal representation of the environment would change at the same timescale that the environment itself changes, rather than at the timescale in which its sensory input changes. Temporal slowness takes advantage of the fact that the many possible visual impressions that can be projected by an individual object make up a single, contiguous manifold in the high dimensional space of visual input (DiCarlo & Cox, 2007). Because of this, different visual impressions of the same object will tend to transform continuously from one to another, tracing a path on that manifold. Therefore visual impressions that occur close to one another in time with high probability will tend to represent two nearby points on the same object's manifold. This information can be used to learn invariant representations in an unsupervised manner, or rather by using time as a supervisor; those input representations that tend to occur adjacently in time are transformed into a common, invariant output representation. This is exactly the kind of learning required by a MHWA such as HMAX, in order to set the connection weights between conjunctive (simple) and disjunctive (complex) cells in each layer.

This principle can be applied in a variety of ways. Foldiak (1991) and Wallis and Rolls (1996) built models of V1 in which the connections between simple and complex cells used a modified Hebbian learning rule that based the magnitude of learning on the product of the current activation strength of the presynaptic simple cell with a measure of the activity over time of the postsynaptic complex cell, using either a short term memory trace or a running average. The result is that the complex cell comes to be associated with a group of simple cells that respond to the same visual input but under various transformations, such as changes in position and scale, that result in input patterns that occur sequentially with high probability in the training data. After training, the model's complex cells exhibited the classical complex response properties of orientation discrimination with phase (position) invariance. Wiskott and Sejnowski (2002) introduced slow feature analysis, which restates the temporal slowness principle as a nonlinear optimization problem, in which, given an input signal, it will find the functions that will produce those output signals that vary most slowly over time, while still carrying significant information. Slow feature analysis produces a set of uncorrelated outputs that vary most slowly in time. When trained with inputs based on V1 simple cells, its results model the receptive field properties of V1 complex cells very closely, including extra-classical properties such as selectivity to direction of movement, end-inhibition, side-inhibition, and variations in orientation and frequency specificity (Berkes, 2005). Unlike the earlier models, however, slow feature analysis remains a mathematical algorithm; there is not yet a biological theory for how it may be implemented by the cortex.

A growing body of evidence indicates that the ventral visual stream uses temporal slowness to learn invariant response properties. Miyashita (1988) trained monkeys on a repeating series of fractal images, and found that some IT cells learned to respond invariantly to several visually distinct images that had appeared consecutively within the training set. Wallis and Bulthoff (2001) presented human subjects with smooth animations of a human head rotating from side to side. As it did so, the face displayed on this head would switch to that of a different person. Subjects would later identify two different faces as belonging to the same individual, if they had occurred in temporal sequence on the same rotating head during training. Cox et al. (2005) were able to 'break' position invariance in human subjects by manipulating the transformation of visual input over time. A visual object would be presented in peripheral vision, and the subject would then saccade to view the object directly. While the saccade was taking place, the object would be swapped out with a similar but visually distinct object. This training regimen resulted in specific confusions suggesting that the images of different objects at various retinal positions came to be associated with a common spatially invariant representation. Li and DiCarlo (2008) confirmed this interpretation by presenting monkeys with a similar object-swap training regimen. Some IT cells developed response properties that were invariant to both multiple positions and multiple object identities, for those cases where an object's identity was switched during saccade. These experimental results support the idea that temporal slowness is used to train the response properties of complex cells in the ventral visual stream, to produce position invariance. Temporal slowness therefore presents a credible solution to the question of how the disjunctive cell properties in a MHWA model such as HMAX may be learned (Masquelier et al., 2007). It also suggests that temporal slowness is instrumental in producing invariance to more complex transformations, such as rotation and pose, for which HMAX has relied on supervised, error-driven learning.

**Predictive Coding**

Another way in which cortical processing may utilize temporal information derived from its inputs is in the generation of predictions. A large body of experimental evidence has shown that, following training with temporal sequences of sensory stimuli, BOLD response is greater when a stimulus violates the temporal pattern established by prior training, than when a stimulus is consistent with that learned pattern. This effect has been seen with natural images (Hupe´ et al., 1998; Bair et al., 2003), auditory streams (Garrido et al., 2007, 2009; Todorovic et al., 2011), apparent motion of visual stimuli (Alink et al., 2010), hierarchical organization of auditory streams (Wacongne et al., 2011), and so on. Colby et al (1996) recorded the responses of parietal neurons during saccade, and found that these neurons first activate in representation of the expected new visual scene immediately before the saccade takes place, and then update to represent the actual new scene once the saccade is complete. Kok et al. (2012) found that prior expectation of a visual stimulus not only reduces the amplitude of the neural response to that stimulus in V1, but also improves the informational quality of its representation.

According to the predictive coding hypothesis (Rao & Ballard, 1999; Srinivasan et al., 1982), a cortical region learns temporal models of its input and uses these models to continuously generate predictions based on its current input in the context of its prior inputs. Prediction errors are then used to update the learned models. The predictive coding hypothesis accounts for the decreased activity due to predicted stimuli, as well as for simple repetition suppression and priming effects (Friston, 2005). Rao and Ballard (1999) introduced a hierarchical model of predictive coding, whereby each level of cortex generates top-down predictions which are compared with novel input at a lower level. Only the difference between the two, the prediction error, is transmitted back up to the higher level. Bastos et al. (2012) have elaborated this into a detailed model of canonical cortical circuits, finding that the computational requirements line up well with anatomical and functional data. De-Wit et al. (2010) note however that it still remains unclear whether predictions are updated at a higher level and passed down to a lower level, or else maintained and updated within a single level. Regardless of the specific mechanisms involved, the predictive coding hypothesis has been rapidly gaining traction as accounting neatly for many different sources of experimental data, and its application has also been extended beyond perception to cognition and mental disorders (Bar, 2009).

In review, there is compelling evidence that the various areas of cerebral cortex all implement a common algorithm. Several lines of evidence point to sparse distributed representation as the medium of encoding used by this algorithm. There are indications that one of the roles performed by this algorithm can be characterized as hierarchical Bayesian inference. Biologically grounded models of visual cortex such as HMAX propose alternating layers of conjunctive units that provide specificity by responding to particular patterns of input, and disjunctive units that provide invariance by responding when any of a number of connected conjunctive units from the previous layer are active. The principle of temporal slowness holds that the underlying external causes of sensory input vary more slowly over time than does the raw sensory input that is projected from those causes, and that meaningful and behaviorally relevant representations can be formed by extracting those more temporally stable statistical regularities from the sensory input. Temporal slowness can be exploited to train the input connections of disjunctive units such as those of the HMAX model in an unsupervised manner. Finally, growing support for the predictive coding hypothesis indicates that the cortex uses temporal information not only for extracting meaningful invariances, but also for recording sequences and making predictions.

**Hierarchical Temporal Memory**

The set of functional constraints described above can be combined with constraints derived from anatomical and other biological data to approach the problem of specifying precisely what processes make up the common cortical algorithm, and how they are implemented by neuronal circuitry. Hierarchical temporal memory (HTM) is a family of learning models and corresponding theories of cortical function that seeks to address this problem by building upon those constraints. HTM is based on Hawkin's (2004) memory-prediction framework, which views the cortex as a bi-directional hierarchy of regions that learn not only co-occurring patterns of input, but also sequences of those patterns. These learned sequences are used to constantly generate predictions of future input, predictions that are passed down to lower levels so as to bias the activity in those levels toward representing input in a way consistent with the ongoing predictions. So long as a region's input is well predicted, that region's feed-forward output remains relatively stable — outputting a pattern that essentially acts as a constant 'name' for the current sequence, with which higher level regions may build further patterns and sequences. In this way, at increasingly higher levels of the hierarchy, patterns of activity will tend to become increasingly stable over time. From this perspective, the spatially invariant complex cells in visual cortex (the disjunctive cells in the HMAX model) are responding predictively; if a particular feature is seen at a given location, there's a high probability that it will be seen at a nearby location soon.

HTM refers to a number of related computational models based on the memory-prediction framework, as well as the cortical theories underlying those models. The first iteration of HTM (George & Hawkins, 2005; George & Hawkins 2009) builds explicitly upon a Bayesian belief propagation learning model, and is made up of representations that correspond to patterns of activity and sequences of patterns within a cortical region. This version of HTM has the strength of being well characterized mathematically, but also has several weaknesses deriving from its abstract level of representation. The second iteration of HTM (Hawkins et al., 2010) is based in much greater detail on its underlying cortical model, with representations corresponding to individual columns, cells, and dendrite segments. While this version does not map as transparently as the first version to the mathematics of Bayesian inference, it gains the advantages of sparse distributed representations (Hinton, 1984; Földiák, 2002) and borrows from several well understood neural network approaches.

In considering the use of HTM as a provisional model of cortical function, it would be useful to review the various iterations of HTM in greater detail and examine their strengths and weaknesses.

**The Zeta 1 Implementation of HTM**

In its first iteration (sometimes called *Zeta 1*), an HTM learning system consists of a hierarchical tree of basic computational units called nodes, each of which is analogous to an area of cortex. The processing that is performed by each node is separated into two stages. The first stage, called the *spatial pooler*, learns and represents spatial patterns: co-occurrences of particular inputs from its child nodes. The second stage, the *temporal pooler*, learns and represents sequences of those spatial patterns. The feed-forward output from a node represents the set of those temporal sequences that are inferred to have the highest probability of representing the node's current input. Because the spatial pooler of each node in the hierarchy learns spatial patterns representing the simultaneous activity of a number of neighboring child nodes, and because the temporal pooler learns sequences of those spatial patterns that occur with high probability, representations at higher levels in the hierarchy will generally represent larger areas of space and longer durations of time than those at lower levels. At higher levels of the hierarchy, activity varies more slowly and is increasingly stable over time. In this way an HTM hierarchy manifests the principle of temporal slowness.

In the Zeta 1 implementation of HTM, learning occurs in each hierarchical level of nodes separately and sequentially. That is, the lowest level of nodes will perform learning first. After being sufficiently trained, nodes at this level will then stop learning and switch to performing inference, while nodes at the second level will begin learning. Because the input to each level of the HTM hierarchy is given in terms of active sequence representations from the immediately lower level of nodes, it is required that the repertoire of nodes at one level be stabilized before learning may begin at the next level. This requirement precludes online learning and is clearly at odds with the performance of the actual cortex.

A node's spatial pooler learns representations of patterns of co-occurring inputs. When performing inference, the representations of the pattern or patterns that most closely match the current input are activated. In its simplest instantiation, the spatial pooler will simply memorize each separate pattern of coinciding inputs that occurs. In real world situations this is usually impractical because it would lead to the creation of too many individual representations. This problem is effectively solved by recording only a fixed number of coincidence patterns, randomly selected from the full set of input patterns, and then allowing more than one pattern to be active at one time, resulting in a sparse distributed activation of coincidence patterns that, as an ensemble, approximates the actual current input pattern.

A node's temporal pooler learns representations of commonly occurring sequences of those coincidence patterns learned by the spatial pooler. Each sequence is recorded as a Markov chain that represents a series of coincident patterns that occur sequentially in time with high probability. A number of different methods have been used to learn the mixture of Markov chains; a simple technique is to learn a transition matrix for the various coincidence patterns and to use a graph partitioning algorithm to identify a set of groups representing the highest probability sequences.

To perform inference, HTM nodes use Bayesian belief propagation (Pearl, 1988). At each time step, a child node passes a message to its parent node indicating the degree of certainty over each of the child node's Markov chains. The parent node uses this information to determine its own likelihood values for each of its coincidence patterns. By making use of the history of messages received (which is collapsed into a state variable that is updated at each time step), the parent node then determines its degree of certainty over each of its Markov chains (ie., it determines which sequences have the highest probability of currently being played out). This information is then passed feed-forward to the node's own parent node. A node also passes back to its child nodes the degree of certainty determined for each of its child node's Markov chains. A node receiving this feedback message then combines it with its feed-forward likelihood values to determine the belief distribution over its own coincidence patterns. In this way bottom-up and top-down information is combined at each level to determine the spatial coincidence patterns and the temporal Markov chains that are believed most likely to correspond to the external causes of the network's input. The belief propagation equations used at each step are examined in detail in (George & Hawkins, 2009).

HTM is an attempt to bridge the gap between a learning framework of spatio-temporal Bayesian belief propagation and cortical function, and so a biologically grounded cortical circuit model is provided for each step of the learning and inference processes. In this model, each cortical column responds to a particular coincidence pattern. It is hypothesized that a column's layer 4 pyramidal cells receive weighted inputs from neurons representing each of that pattern's constituents at the lower level. Columns mutually inhibit their neighbors, and this competition sharpens the selectivity of the pattern learned by each column. This type of competitive Hebbian learning process is typical of many connectionist models of cortex that perform feature detection or vector quantization (e.g., Kohonen, 1982; Rumelhart & Zipser, 1985; Churchland & Sejnowski, 1992; Tsunoda et al., 2001; Buzo et al, 1980).

It is its temporal learning properties that set HTM apart from other models. Within a column are also a second set of cells (posited to be layer 2/3 pyramidal cells) representing the column's coincidence pattern in the context of each of the various sequences (Markov chains) in which it may occur. These cells receive lateral input from cells in other columns, which are active in response to other spatial patterns that are part of the same Markov chain. These contextual cells are driven by the weighted sum of this lateral input multiplied by the input from the column's cells that are active in response to the column's spatial pattern being active in any context. Thus each of these contextual cells is activated only when its column's spatial pattern is active within the temporal context represented by a single Markov chain.

A third type of neuron (also hypothesized to be located in layers 2/3) receives input from all cells that belong to a single Markov chain, regardless of which spatial pattern within that Markov chain's sequence that they represent. These cells are active whenever any spatial pattern within the chain (any step within the sequence) is active. By pooling the output of all cells that represent any step within a single Markov chain, these cells perform the role of the disjunctive layer in an HMAX network. It is these cells that send their output feed-forward to the next higher layer (the parent node) in the network. Two additional neuron types (located in layers 5 and 6) are hypothesized to handle the functions of top-down message passing. The biological evidence that provides justifications for the hypothesized mappings of HTM's belief propagation equations to cortical circuits is reviewed in (George & Hawkins, 2009), along with a set of predictions derived from the model and an examination of possible variations.

The Zeta 1 version of HTM has several fundamental weaknesses (George & Hawkins, 2009; Maltoni 2011). The first of these weaknesses is in the rigidity of its representations. A Zeta 1 node's spatial representations consist of discrete coincidence patterns that each record a specific set of co-occurring inputs. Each such coincidence pattern represents the entire state of the input to that node. This presents several problems: every meaningful variation of input state would need to be represented by a separate coincidence pattern, resulting in the generation of very large numbers of coincidence patterns when given rich input data; there is no means of generalization that would allow learning associated with one input pattern to be utilized by a similar second input pattern; and there is no support for compositionality, whereby multiple separate subpatterns of input may be active at one time. Several partial solutions to these problems were implemented in Zeta 1. The profusion of coincidence patterns that would result from storing every individual input pattern was addressed by storing only a fixed number of different coincidence patterns, and activating the pattern that most closely matches the input. The lack of support for compositionality and generalization was in part addressed by allowing multiple coincidence patterns, each being a partial match to the current input, to be active at one time, resulting in a sparse distributed representation that would approximate the actual input. Topologically organized compositionality, such as the activation of many different patterns representing individual localized features in V1, was addressed by splitting the representation of a single cortical area into many separate HTM nodes with overlapping receptive fields.

Related to the problem of the rigidity of its representations, the Zeta 1 implementation is incapable of online learning. Learning must begin at the lowest level of the hierarchy, where the spatial representations (coincidence patterns) are learned first, followed by the temporal representations (Markov chains) that build upon those spatial representations. Learning then proceeds for the next higher hierarchical level, and so on. Because each level of representation is built in terms of the representations at the previous level, learning must cease at one level so that its representations become stable, before learning can begin at the next level. An HTM network can only begin to be used for inference after learning has been completed at all levels in series.

A third important weakness in the Zeta 1 version of HTM is with the temporal properties of the method in which a Markov chain's probability is determined during inference. Originally (George, 2008), the probability of a Markov chain was based only on the immediate probabilities determined for the spatial coincidence patterns at the current time step, with no regard given to the sequence of prior states. This method would result in first order predictions based only on current activity, which works well for modeling spatial invariance but does not successfully model temporal sequences except in the simplest case where each coincidence pattern takes part in only a single temporal sequence. This system was later modified (George & Hawkins, 2009) to take into account the sequence of previously active coincidence patterns (as well as the currently active coincidence patterns) when determining the probability for each Markov chain. This system was capable of both modeling spatial invariance and of making nth-order sequence predictions based on prior context, but it could only represent sequences that advance at a constant speed, one step per time tick.

The Zeta 1 implementation also does not address some important aspects of cortical function at all. Top-down feedback is used only to implement the Bayesian belief propagation equations, and does not affect a node's feedforward output. Therefore it is insufficient to model the role of feedback in selective attention. Finally there is no attempt to model the motor output function of cortical layer 5, with the timing and gating functions this would require.

**The HTM Cortical Learning Algorithms**

Many of the problems discussed above were addressed in the second iteration of HTM, called the cortical learning algorithms (CLA). The CLA is based on a model of cortical function similar to Zeta 1's, and at a high level it embodies most of the same learning principles as Zeta 1. However, the CLA employs entirely different algorithms and representations, implementing its model of cortical function at a more granular and biologically detailed level of abstraction.

The basic units of representation in the CLA correspond to cortical regions, cortical minicolumns, cells within those columns, and dendritic branches. Following the biological model, each column is considered to represent, and to activate most strongly and selectively to, a particular subpattern of inputs. As a simplification of biological detail, each column is given a single dendrite segment through which it receives input from lower hierarchical levels. This is termed a *proximal* segment, because it corresponds to a layer 4 neuron's dendrites that are close to the cell body, where they receive feedforward input from axons projected from neurons in lower cortical regions. A column's proximal segment receives input from a random subset of all inputs that project to its cortical region, and that subset may be most dense centrally and fall off according to a Gaussian function, in order to support topologically mapped inputs.

At any given time only a sparse subset (e.g. 2%) of columns within a region will be active. The spatial pooler stage of the CLA will determine the degree to which the subpattern of input represented by each column's proximal dendrite matches the current input. Only those within a local area that best match their active input will be activated, by way of a k-winners-take-all (kWTA) process intended to model local lateral inhibition within a cortical region. Columns which have been least active over time are "boosted" to be more sensitive to their inputs and so more strongly competitive with other columns, in order to encourage the participation of all columns within the set of representations and to enable the reuse of columns representing input patterns that are no longer common. After the inhibition stage, when the final set of active columns has been determined, the proximal synapses of only those columns that are active have their synaptic strengths modified according to a Hebbian learning rule. The result is that, over the course of learning, the activity of each column comes to represent a specific combination of inputs that occur with high probability. The CLA's spatial pooler thus implements a form of competitive learning (Rumelhart & Zipser, 1986), akin to conditional principal component analysis (O'Reilly & Munakata, 2000) or Kohonen's (1982) self-organizing map. In the CLA, columns fill the role of Zeta 1's coincidence patterns, by representing individual co-occurrences of inputs. The CLA spatial pooler's learning algorithm solves the problem of choosing which limited set of input patterns to represent by using this well understood method to converge toward a set of most common subpatterns. The use of a sparse distributed representation where each column represents only a small sample of the region's full input allows a complete pattern of active inputs to be represented approximately in a way that supports generalization and compositionality.

The CLA's biologically inspired sparse distributed representation of inputs offers many advantages over Zeta 1's set of discrete coincidence patterns. However it also introduces complexities. Temporal sequences can no longer be represented as Markov chains of spatial coincidence patterns, because those spatial patterns are now composed of a sparse compositional coding in which a given exact set of columns may rarely if ever be active more than one time. Instead, the temporal pooler of the CLA takes a very different approach.

The CLA and Zeta 1 are based on similar theories of cortical function. This theory holds that while each cortical column represents a single specific spatial pattern of inputs, an individual cell within that column will respond only to that spatial pattern in the temporal context of a particular pattern of prior activity within the same cortical region.

While Zeta 1 abstracted this model, the CLA implements it explicitly. In the CLA, a cell may be inactive, or it may be in one of two states of activity. It may be *active*, meaning that it is currently representing feedforward input, or it may be *predictive*, meaning that it is primed or expecting to be actively representing input soon. In (Hawkins et al., 2010) it is hypothesized that the active state corresponds to bursting activity of a cortical pyramidal neuron, whereas the predictive state corresponds to tonic firing. (Another possible neurophysiological basis for the predictive state, suggested by Hawkins (personal communication, May 13, 2013), is with a modified Hebbian synaptic modification rule with a short term memory trace, as described by (Wallis & Rolls, 1997).) A cell is put into predictive state by receiving above-threshold excitatory input via a *distal* dendrite segment. Distal segments project horizontally within layer 1, where they receive input from cells in the same cortical region. An individual synaptic depolarization on a distal dendrite has very little effect by the time it reaches the soma, however multiple simultaneous synaptic depolarizations on the same segment of a distal dendrite can trigger the firing of the receiving cell (Mel, 1999). This nonlinear property of distal dendrites allows individual dendritic segments to act as separate coincidence detectors that respond to a particular pattern of activity of nearby cells within the same cortical region. In the CLA, a cell's distal segments learn a set of patterns of activity of nearby cells within the same region (by learning to respond to sub-samples of those patterns), which have previously led to the subsequent activation of the cell. When such a pattern of local activity is later detected, the cell is put into predictive state.

When a column is activated, representing the present activity of a particular pattern of feedforward inputs, then if any of the cells within that column were already in predictive state, only those cells transition to active state. All other cells in the column remain inactive. The set of active cells in the column therefore represents the current activity of the spatial pattern to which the entire column responds, but within the temporal context of prior local activity represented by the particular cell(s) that were predicted to activate (ie., were in predictive state). In this way only a subset of a column's cells are activated when it is part of a predicted sequence, the subset representing activity within the temporal context of that sequence (Figure 1). If none of the cells in the column are in predictive state when the column is activated, then all of the cells in the column transition to active state. The activity of all of the cells represents the column's spatial pattern without preference to any particular contextual interpretation, by activating the cells that represent all possible interpretations. This approach is consistent with findings that prior expectation of a stimulus reduces neuronal activity when the stimulus occurs, which is the basis of the predictive coding hypothesis.

A cell's distal dendrite segments learn these predictive associations with neighboring cells using a modified form of Hebbian learning. A cell $c$ that is active at time $t$ has those distal synapses that are connected with cells that were active at time $t$-1 strengthened. The result is that in the future, if a pattern similar to the active pattern at time $t$-1 occurs, there will be an increased likelihood of cell $c$ transitioning to predictive state. In addition, a cell $c$ that is predictive at time $t$ has those distal synapses that are connected with cells that were active at either time $t$ or time $t$-1 modified. If cell $c$ later transitions from predictive state to active before becoming inactive (meaning that its predictive state represented a prediction that in this case was true), then those synapses are strengthened. Otherwise, cell $c$'s predictive state represented an incorrect prediction, and so those synapses are weakened.

It is the combination of a region's active and predictive cells that send feedforward signals to the next higher level in the CLA hierarchy. Depending on the frequency and reliability of repetition of a given sequence, a CLA region may come to predict some or all of the continuation of a sequence once it has received enough context to identify that specific sequence. Because of this, and because a CLA region outputs signals from both active and predictive cells, the output of a region will become more stable over time than its input, and particular static patterns of output will come to correspond with particular sequences of input. In this way, the output of a CLA region has similar properties to the set of active Markov chains that was given as the output from a Zeta 1 node; both implementations take advantage of the principle of temporal slowness. However, the CLA's temporal pooler is much more flexible than Zeta 1's. It is compatible with CLA's sparse representations of spatial input patterns, and propagates the properties of sparse coding, including generalization, compositionality and efficiency of storage, to the representation of sequences. It is capable of continuous online learning, given that the learning rate is sufficiently slow, because the sparse representations at each level will gradually adapt as the lower level representations that they are built upon change. Finally, the CLA's sequence representations are sensitive to the specific order of the elements of a sequence, but are not bound to specific timings of sequence elements as were the Zeta 1 sequence representations. So long as a sequence's elements are activated in the correct order, the cells representing the elements of that sequence will become predictive and then active. This allows the CLA to respond invariantly to a given sequence presented at various rates of speed.

The theory of cortical function underlying the CLA version of HTM is very similar to that of the Zeta 1 version. One difference is in the proposal of two different activity states for a pyramidal cell, active (due to feedforward excitation via proximal dendrites) and predictive (due to lateral excitation via distal dendrites). Another difference is that the CLA implementation does not yet incorporate top-down feedback connectivity in any way, and so it remains agnostic on the issue of the several types of neuron that were proposed as part of the Zeta 1 theory in order to map a neuronal equivalent to the feedback portion of the Bayesian belief propagation equations.

The CLA algorithm as described learns nth-order sequences, incorporating variable length temporal context and making specific predictions of subsequent activity based on the current input combined with that context. This results in the learning of specific sequences, and is considered to correspond most closely with the function of cortical layers 2/3. In V1, it is proposed that the response properties of cells that are selective to both orientation and direction of motion are learned in this way; such cells are found in layers 2/3 (Hirsch & Martinez, 2006). However, this system does not explain how the response properties of V1's complex cells may be learned, so as to be selective to a particular orientation with spatial invariance. Representing spatial invariance would require predictively activating all columns that respond to a particular orientation that are local to a column that is activated feedforward by input matching that orientation. This means ignoring prior context (direction of movement) and predicting based solely on the current input. One way to drop temporal context information in the CLA framework is to remove inhibition between the cells in a column that represent a particular input in various specific temporal contexts, so that when any of them are activated, they will all become active. It is proposed that while layers 2/3 learn specific sequences and make predictions based on temporal context, layer 4 learns only first order sequences, making predictions based only on current feedforward input. This would satisfy the requirements for learning spatial invariance by taking advantage of the temporal slowness principle (Figure 2). This effect can be obtained using the CLA algorithm by simply using a single cell per column. The proposal that first order sequences are learned by layer 4, while speculative, is supported by the presence of complex cells in layer 4 (Martinez et al., 2005), as well as the finding that layer 4 is thickest within sensory areas and thinner in association cortex, where presumably spatial invariance information has already been fully extracted (Hawkings et al., 2010).

One aspect of cortical function that is not addressed by the CLA version of HTM is the encoding of the specific timing of sequence elements. Whereas the Zeta 1 implementation recorded a sequence as a series of events attached to specific time steps, the CLA encodes the order of sequence elements but does not specify timing. This is useful for generalization and mirrors some aspects of biological performance. However, many studies have demonstrated that the brain is capable of timing discriminations at the scale of hundreds of milliseconds, and that this ability is crucial for such tasks as coordinated movement and speech sound discrimination. Theories differ as to whether the source of the timing signal is centralized or distributed (Buonomano & Karmarkar, 2002), but it is known to be represented by the time-variant activity of neuronal populations throughout the cortex (e.g., Jin et al., 2009; Schneider & Ghose, 2012). It is proposed in (Hawkins et al., 2010) that specific timing is learned by cortical layer 5, using the same CLA mechanism used to learn sequences in layers 2/3, but incorporating a timing signal (possibly received from elsewhere in the brain via layer 5's thalamic afferent, see (Rodriguez et al., 2005)) in order to allow layer 5 cells to become predictive only for the very next sequence element, and only at the time at which its activity is expected. Layer 5 is seen as being the most likely candidate for this role because it is the source of the cortex's motor output signals, as well as of feedforward output that is gated through the thalamus.

The CLA version of HTM is a significant improvement over Zeta 1. Its use of sparse distributed representations for encoding both spatial and temporal patterns allows for generalization, compositionality and online learning. Its method of encoding sequences preserves sequence order without requiring exact timing. And its biologically detailed level of abstraction makes it straightforward to extend and modify to incorporate additional aspects of cortical function.

The CLA is not without its own weaknesses, however. As a biologically based model its components don't map directly to the equations of Bayesian belief propagation, as those of Zeta 1 did. This means that while the CLA is capable of similar processing as Zeta 1 but with greater flexibility, it is also more difficult to characterize mathematically. The CLA currently lacks any kind of top-down feedback, which would be required for a complete implementation of Bayesian belief propagation (which Zeta 1 had) as well as for additional cortical functions such as selective attention. The current CLA implementation also lacks the ability to discriminate sequences based on the specific timing of their elements, as discussed above. Compared with Zeta 1, the CLA has been utilized less outside of the company Grok (where both versions of HTM were developed; formerly known as Numenta), and has received little attention in academic publications (e.g., Price 2011; Thornton et al., 2012; Zhou & Luo, 2013). Finally, for reasons of processing speed and memory efficiency the CLA makes several simplifications of biological details that most connectionist learning models do not, such as binary cell activation states (as opposed to rate coding) and binary synaptic strengths controlled by linear weight values. These may prove to be suboptimal for some applications.

The HTM learning model elegantly combines a number of key capabilities attributed to cortex, including the hierarchical dimensionality reduction of input using sparse distributed representation, the unsupervised learning of invariance using temporal slowness, and predictive coding, and its corresponding theory of cortical function maps well to a large set of functional and anatomical data. HTM is by no means a complete model of cortical function, nor is it yet fully successful in solving those problems which it does address. But because of its unusual set of strengths, I believe that further study, application, extension and refinement of the HTM CLA is warranted. To date, applications of HTM have focused almost entirely on pattern classification problems typical of posterior cortex. If the uniform structure of the cerebral cortex does point to a universal cortical algorithm, and if HTM (or some elaboration of it) is to be considered a candidate model for this fundamental cortical function, then HTM should be able to be applied as successfully to prefrontal and motor cortical function as it is to posterior cortical function. In the interest of expanding the application of HTM to a wider range of cortical functions, I will next review several leading models of frontal cortical function, and look at the degree to which a theory of cortical function with the properties of HTM would be compatible with these models. Through such a comparison, it may be possible to expose some areas where HTM is well suited to be applied to frontal cortical functions, and other areas where the requirements of frontal cortical functions may indicate ways in which HTM may be improved. Likewise, this exercise may suggest ways in which models of frontal cortical function may benefit from being paired with a cortical learning algorithm having the properties of HTM.

## The Frontal Cortex

The frontal lobes have long been among the least well understood regions of the brain. Early experiments using lesion and electrical stimulation failed to identify any clear functions for the majority of frontal cortex, resulting in their being considered the "silent lobes" for much of the twentieth century. It is only in the last several decades that new techniques have allowed researchers to begin to build an understanding of frontal lobe function.

The frontal lobes are typically subdivided into two functionally distinct regions: the motor areas located at the caudal end, adjacent to the central sulcus, and the prefrontal cortex (PFC) located at the  rostral end, behind the forehead. Ferrier (1874) first mapped the motor areas using direct electrical stimulation. He found that stimulation of specific areas of motor cortex would produce movement in corresponding muscle groups, resulting in a rough somatotopic map. Brief stimulation would produce muscle twitches, while more prolonged stimulation would produce coordinated sequences of seemingly meaningful movements, such as stepping or reaching. The motor areas are further subdivided into the primary motor area, the premotor area, and the supplementary motor area, based on cytoarchitectonic and cell response properties. Together, these areas make up the origins of the majority of axons projecting from the brain to the spinal cord via the cortico-spinal tract, for control of voluntary movement.

### Functions of the Prefrontal Cortex

It is the prefrontal cortex that has only begun to yield its secrets in the last few decades, particularly with the use of functional imaging and single unit recording methods. Functional roles that are known to depend primarily upon the prefrontal cortex include the representation of task, goal, strategy and reward; strategy and response selection and initiation; the inhibition of a prepotent response in favor of a weaker, more appropriate response; top-down direction of attention; active maintenance of working memory representations; and the storage and retrieval of explicit memories. Critical evidence for the association of prefrontal cortex with each of these functional roles will be summarized below.

**Representation of task, goal, strategy and reward.** The learning of arbitrary conditional associations between cue and response is disrupted by PFC damage in both monkeys (Halsband & Passingham, 1985; Petrides, 1985) and humans (Petrides, 1990). Furthermore, the pattern of activation in the PFC has been shown to correspond to the specific task rule that is guiding behavior. Assad et al (2000) trained monkeys on a task that employed multiple rules but used the same cues and responses across the different rules. The response properties of more than half of lateral PFC neurons were found to be rule dependant.

Baker et al (1996) showed that the prefrontal cortex is active in normal subjects when performing the Tower of London task, which requires planning several moves ahead. Shallice (1982) found that prefrontal patients are significantly impaired when performing this kind of task, making arbitrary moves that are not guided by a larger plan.

Level of activation in a region of orbital PFC has been found to correspond with the reward value of a specific stimulus (O'Doherty et al, 2000), and damage to ventromedial areas can lead to sociopathic behavior and difficulty making appropriate choices (Price, 1999).

**Strategy and response selection and initiation.** Patients with prefrontal damage will often manifest behavior that is composed of fragmented action sequences which leave out relevant steps while including other, irrelevant actions. Well practiced behaviors are maintained, but the ability to actively guide behavior toward a purpose is diminished (Duncan, 1986). Prefrontal damage also leads to behavior that is impulsive and lacking in regard for consequences (Luria, 1969).

Another hallmark of prefrontal damage is the perseveration of a strategy when it is no longer successful. For example, prefrontal patients have increased difficulty with the Stroop task when the response rule changes frequently (Dunbar & Sussman, 1995; Cohen et al, 1999).

**Inhibition of prepotent response.** The prefrontal cortex is instrumental in selecting responses that are appropriate not only to immediate environmental cues but also to latent elements of context such as goals and social norms. Patients with prefrontal damage will often engage in "utilization behavior" (Lhermitte, 1983), making use of whatever utilitarian object is present in the environment regardless of its relevance or appropriateness in the current context.

In the Stroop task, subjects are required to override a stronger, more well practiced action selection rule that is task-irrelevant (reading) with a weaker, more unusual rule that is task-relevant (color naming). Patients with compromised prefrontal function have a specific difficulty with overriding the prepotent response rule when performing this task (Perrett, 1974; Vendrell et al, 1995). Similarly, patients with prefrontal lesions have difficulty overriding the prepotent response of saccading toward a stimulus, in a task in which the objective is to saccade in the opposite direction (Guitton, Buchtel & Douglas, 1985). Humans and animals with impaired prefrontal function have particular difficulty with tasks such as the Wisconsin Card Sorting Test (WCST) and its analogs, which test flexible adaptation to changing task demands (Milner 1963, Dias et al 1996, Rossi et al 1999).

Prefrontal cortex is also implicated in verbal fluency, and prefrontal damage that disrupts verbal fluency as measured by a standard sentence completion task is separable from other loci of prefrontal damage that impair performance on a nonsense sentence completion task, in which the object is to complete a sentence in a way that doesn't make sense. (Shallice & Burgess, 1996). Standard sentence completion can employ well-learned response patterns, while nonsense sentence completion requires more complex self-organization, and inhibition of the prepotent response.

**Top-down direction of attention.** PFC activity corresponds with the magnitude of effort required to maintain directed attention. During a task that requires attending to a particular stimulus dimension, prefrontal activity is increased when other stimulus dimensions conflict competitively with the target dimension (Banich et al. 2000).

In a study described by Shallice and Burgess (1991), prefrontal patients with normal IQ scores failed to carry out simple tasks for which they were given explicit instructions (such as shopping for several items) because their focus was distracted by intervening events.

Visual attention also suffers with impaired prefrontal function. Gaze shifts become haphazard, halting, and prone to repetition (Luria, 1966; Tyler, 1969). One region of the PFC, called the frontal eye field (FEF) due to its association with visual attention, is activated in a wide range of visuospatial tasks (e.g., Corbetta et al, 1993, 1998; Fink et al, 1997; Kastner et al, 1998), and cell recording studies have shown activity in the FEF corresponding with both saccades (Wurtz & Mohler, 1976) and with covert shifts in visual attention (Kodaka et al, 1997). The FEF projects feedback connections to both the ventral visual cortical areas and to the parietal cortex, making it ideally suited for the direction of both feature based and location based visual attention (Kastner & Ungerleider, 2000).

**Active maintenance of working memory representations.** Many studies have focused on the prolonged transient activity seen in prefrontal neurons during the delay period of a delayed-response task. In some cases the pattern of delay period activity was found to correlate with the specific cue dimension (such as identity, color, or location) that would need to be remembered during the delay period in order to successfully guide behavior (e.g., Fuster, 1973; Fuster, 1982). In other cases, delay period activity was found to correlate with impending action (Asaad et al, 1998) or expected reward type (Watanabe, 1996) or magnitude (Leon & Shadlen, 1999). This maintained activity was also robust to interference by distracter stimuli presented during the delay period (Miller et al., 1996). Conversely, impaired PFC function results in increased distractibility (Chao & Knight, 1997).

**Storage and retrieval of memory.** A variety of studies implicate the prefrontal cortex as having an important role in the retrieval of explicit memories (Yener & Zaffos, 1999). The PFC is reciprocally connected with the hippocampus and closely related regions (Leichnetz & Astruc, 1975; Rosene & van Hoesen, 1977), and lesions to dorsolateral PFC typically result in impaired explicit memory retrieval (Schmaltz & Isaacson, 1968; Wikmark et al, 1973).

Janowsky et al (1989) conducted explicit memory tests of patients with prefrontal lesions and found that they were impaired in free recall, which depends on self-organization and retrieval strategies, despite good performance in recognition. Gershberg and Shimamura (1995) found that frontal patients were impaired in the use of strategies during both the encoding and free recall of explicit memories, resulting in compromised performance. Schacter (1997) found increased prefrontal blood flow associated with retrieval of episodic memories.

**Functional Specializations of Frontal Regions**

The frontal cortex is composed of a number of areas that can be differentiated by their cytoarchitecture and connectivity. There is also a growing body of knowledge about how these areas differ functionally.

The most basic distinction among frontal areas is between the motor areas and the prefrontal cortex. The motor areas consist of primary motor cortex (Brodmann area 4) and the premotor and supplementary motor areas (Brodmann area 6). These areas are agranular (lacking in a well defined layer 4) and typically produce muscle movement when electrically stimulated. Adjacent to and connected with the premotor area is the dorsolateral prefrontal cortex (Brodmann areas 8, 9 and 46) and the ventrolateral PFC (Brodmann areas 44, 45 and part of 47). At the anterior pole of the brain is the orbital frontal cortex (Brodmann areas 10, 11, and part of 47). The prefrontal areas are granular (or dysgranular in the case of dorsolateral area 8, which is transitional between the granular PFC and the agranular premotor area), and do not typically produce muscle movement when electrically stimulated.

The lateral areas of the prefrontal cortex have been implicated in the learning of stimulus-response rules, and in the inhibition of well-learned rules in favor of weakly-learned rules that are more task appropriate given the current context (Diamond & Goldman-Rakic, 1989). Lesions to the lateral areas compromise both of these functions (e.g., Milner, 1963; Bussey et al, 2002). Sustained activity in seen in the lateral PFC during a delay task; this activity typically corresponds with a cue stimulus that is being remembered in order to guide response after the delay, with the specific rule being employed to guide a response, or with a specific action being planned (Constantinidis et al, 2001; Assad et al, 2000; D'Esposito et al, 2000). The lateral PFC is reciprocally connected with higher level sensory areas in posterior cortex, as well as with premotor cortex, and so it is well placed to build representations that connect perception with action (Fuster, 1997).

As a stimulus-response rule becomes well learned, however, the role of the lateral PFC in support of that rule decreases. Well learned rules are robust to lateral PFC lesions, and Wallis and Miller (2003) showed that premotor cortex response distinguishes between well-learned rules more than 100ms earlier than does lateral PFC response. This indicates that as a rule is trained extensively, it is learned directly by premotor cortex and no longer requires the support of lateral PFC in order to produce behavior that is guided by that rule (Bunge, 2004).

The medial prefrontal cortex likewise receives sensory information from posterior areas, but unlike the lateral PFC it is also connected with brain centers primarily involved in emotion and motivation, such as the amygdala and the cingulate cortex. The medial PFC contributes to conflict monitoring (Botvinick et al, 2001), decision making under conditions of uncertainty (Rushworth & Behrens, 2008), outcome evaluation (Gehring & Willoughby, 2002), social interaction (Amodio & Frith, 2006), and the processing of emotion (Etkin et al., 2006). While lateral PFC specializes in the execution of cognitive and behavioral actions, medial PFC specializes in emotion, motivation and decision making.

The most posterior area of medial PFC (adjacent to the premotor area) is concerned with the selection of individual behaviors, activating strongly when there is conflict between competing responses. Further anterior to this is an area of medial PFC that is concerned with decision making, with activity corresponding to the relative desirability of multiple decision options. Further anterior still is an area that is sensitive to strategy, which activates preferentially in response to deviation from the subject's chosen decision making strategy (Venkatraman et al, 2009). A similar pattern of activity ranging from concrete to abstract representations was found in lateral PFC by Christoff et al (2009). These results indicate a hierarchical organization of control in the frontal cortex, from the level of concrete movements in the motor and premotor areas, through specific behaviors in posterior PFC, to more abstract concepts and strategies in anterior PFC.

The posterior cortex is frequently characterized as having two distinct processing streams. The dorsal or 'how' stream (involving the occipital and parietal lobes) is concerned with properties of the physical environment that are relevant to the guidance of action, while the ventral or 'what' stream (involving the occipital and temporal lobes) is concerned with the identification of objects in the environment. There is evidence that this system of organization extends to the frontal lobe as well. Anatomically, ventral PFC is connected primarily with ventral areas of posterior cortex, and dorsal PFC is likewise mainly connected with dorsal posterior cortex. While activity in much of the PFC shows correspondence with a mix of stimulus ('what') and response ('how') information, ventral and dorsal PFC do appear to have some specialization for 'what' and 'how' processing, respectively; Nagel et al (2008) found that manipulations of a task's response selection difficulty would modulate activity in dorsolateral PFC, while manipulation of semantic selection difficulty would modulate activity in the ventrolateral PFC.

In summary, there is evidence for functional specialization in the prefrontal cortex along three axes: medial areas are concerned with 'hot' value and emotion based processing while lateral areas are involved in 'cold' cognitive and motor processing; posterior areas are concerned with concrete actions while anterior areas process abstract concepts and strategies; and ventral areas are part of the brain's 'what' object identity stream while dorsal areas are more concerned with the 'how' of action guidance (O'Reilly, 2010).

**Models of Frontal Cortical Function**

So far I have reviewed a variety of functions attributed to frontal cortex, as well as some patterns that underlie the localization of these functions. The distinguishing functions of frontal cortex, including working memory, attentional set, and the representation and selection of actions and strategies, would appear to have little in common with the characteristics of posterior cortex as understood through a model such as HTM — the representation of sensory input with increasing complexity, temporal extension, and invariance at each successive hierarchical level. As a first step toward reconciling these differences, it would be useful to review a number of different models that propose specific mechanisms underlying frontal cortical function. Most of these models are driven by a particular subset of frontal phenomena, but taken together they begin to build a coherent picture that is compatible with both biological and cognitive constraints.

**Baddeley's working memory model.** Building upon Atkinson and Shiffrin's (1968) multi-store model, Baddeley introduced an influential model of executive function based on multiple working memory stores (Baddeley & Hitch, 1974; Baddeley, 1986).

Motivated by experiments showing that subjects could perform two working memory tasks simultaneously without degraded performance only if they did not share the same perceptual domain, Baddeley's model consists of two domain-specific memory store "slave" systems, the phonological loop for written and spoken language and the visuo-spatial sketch pad for visual information and navigation, as well as a central executive to control these components. A later update of the model (Baddeley, 2000) added a fourth subsystem, an episodic memory buffer, to help explain the interactions between the central executive, working memory, and long term memory.

Innovations provided by Baddeley's model included the separation of the previously unitary concept of short term memory into multiple modality-specific memory stores, as well as emphasizing the active, "working" aspect of short term memory, as embodied by the central executive. The central executive is seen as being in charge of attention, action and cognition, of selecting what is stored in short term memory, and of controlling interactions with long term memory. It is presented as a unitary concept, and Baddeley develops no theory for how it carries out its functions.

**Attentional control.** Norman & Shallice (1980, 1986; Shallice, 1982; Shallice & Burgess, 1991, 1996) proposed a model of executive function that is rooted in the concept of a production system (e.g. Newell, 1973; Anderson, 1983). A schema, consisting of a sequence of cognitive or behavioral actions, is activated when a particular trigger pattern is detected. Such a trigger pattern may be made up of perceptual inputs as well as the output states of previously activated schemas. If multiple connected schemas are triggered to be active simultaneously, mutual inhibition prevents all but the most strongly activated schema from being initiated. The more frequently that a schema's trigger leads to the activation of that schema, the more well-learned it becomes, and the more strongly it will inhibit connected schemas. This system, called the contention scheduler, is considered sufficient to handle routine responses and cognitive processes.

For novel or non-routine processes a slower, more flexible system is employed, called the supervisory attentional system (SAS). This system is employed when existing schemas are not sufficient to address the current situation, and it utilizes general, abstract patterns to generate new schemas specific to the situation. The SAS guides the activity of the contention scheduler, when necessary, by providing biasing signals that alter the probabilities of activation for the various schemas.

The SAS is considered to be located in the prefrontal cortex, and is responsible for the functions attributed to PFC such as top-down selective attention, the integration of action and perception over time, working memory, episodic memory retrieval, error monitoring, and the inhibition of automatic responses in favor of more contextually appropriate responses. The functions of the contention scheduler map to those functions considered to be carried out by the frontal cortex (particularly its posterior areas) in cooperation with posterior cortex.

Beyond simply attaching new labels to groups of brain functions, the Norman & Shallice model places focus on the kinds of representations that would need to exist in order for the prefrontal cortex to carry out its work, the process of generating those representations, and the conditions under which the deliberate, supervisory functions of the PFC would be set in motion. Other contributions of this model include the emphasis on a separation between fast, routine, automatic processes and slow, deliberate and flexible processes, and the proposal that the SAS mainly influences the contention scheduler by biasing particular schemas to be more or less likely to activate.

**Temporal integration.** Fuster (1980/1997, 1999) views the frontal cortex as a motor/executive hierarchy with the purpose of ordering sequences of action toward goals. The motor area represents and controls individual movements; premotor representations introduce more abstract properties such as trajectory, and can be agent-invariant as in the case of mirror neurons that fire both when a particular action is executed or is only perceived being performed; prefrontal cortex integrates perception and action to control novel behavior, with representations increasing in complexity and abstraction toward the anterior pole.

Fuster focuses on both the processes carried out by the PFC as well as the representations involved. In his framework, PFC representations are schemas, or temporal gestalts, combining elements that may be very abstract – in some cases defined only by their relationships with one another. These schemas are formed by repeated practice or reenactment of similar actions and situations, such that networks form representing the common elements of these sequences. These networks may then act as action symbols, in the same way that invariant representations in high level posterior cortex act as perceptual symbols. Further abstraction at higher levels may lead to general concepts of action, such as responsibility, altruism, or rule of law.

In Fuster's view, the successful temporal integration of action and perception over time, which forms the foundation for cognition, behavior, and language, is based on the capacity to mediate cross-temporal contingencies, such as "If now this, then later that; if earlier that, then now this". This in turn relies primarily on 3 subfunctions of prefrontal cortex: short-term motor memory or preparatory set, short-term perceptual memory for retaining sensory information upon which future action will be based, and inhibitory control of interference.

As with Norman & Shallice's production system based model, Fuster believes that the execution of a behavior is triggered and maintained by the recognition of a particular pattern of inputs, both from the environment and from internal representations. The result is a feedback loop, with perception influencing behavior, and behavior affecting the environment, which then modifies perception.

**Goldman-Rakic' working memory model.** The model proposed by Goldman-Rakic et al (1987, 1996) is built on her pioneering work studying the active maintenance of frontal lobe representations during delay tasks. While other contemporary models proposed that prefrontal functions such as attention, affect, inhibitory control, motor planning and spatial working memory are mapped to different cytoarchitectonic regions of PFC (e.g., Fuster, 1980; Pribram, 1987), Goldman-Rakic contended that working memory is a common process shared by the various areas of PFC. Rather than different functions being mapped to different PFC areas, each PFC area works with different domains of information, and applies the common process of working memory to each of them.

For example, electrode recording studies in monkeys and human functional imaging experiments show that certain more dorsal PFC areas are active when the subject remembers spatial location information during a delay, while specific ventral PFC areas are active when the subject remembers object and feature information. These two areas are connected to posterior cortical regions appropriate to the domains of information that they handle (parietal and inferotemporal cortex, respectively).

Within a particular area of PFC, Goldman-Rakic proposes that the activity of cells within a single cortical column codes for specific content (for example, a specific location, object, color, or visual feature). Different cells within such a column may respond under different conditions, however; some may respond when a stimulus is registered, others while the representation is being actively maintained during a delay period, and still others correspond with response preparedness. Goldman-Rakic stresses that these three categories of sensory, memorial and motor subfunctions are represented within  the same microarchitectural cortical module, rather than being compartmentalized into separate PFC regions.

The executive functions of the PFC had previously been considered to be governed by a polymodal and general purpose mechanism in charge of control and selection processes, for example Baddeley's (1974) 'central executive' or Norman & Shallice's (1980) 'supervisory attentional system'. Goldman-Rakic's model took a first step toward decomposing this "homunculus"-like executive system by proposing that executive processing is the result of interactions between multiple parallel, independent, domain-specific processing modules that each incorporate sensory, working memory, and motor control functions, and that each utilize connections with domain-appropriate posterior, premotor and limbic brain regions.

Another important contribution of the Goldman-Rakic model was to propose that the prefrontal function of inhibiting prepotent responses in favor of more context appropriate responses, which when compromised results in perseveration and distractibility, does not reflect an independent and localizable 'inhibitory' function of the PFC. Instead, prefrontal working memory provides the activation bias necessary to select a weaker correct response over a prepotent incorrect response; disturbance of this working memory function removes that bias, allowing the default, prepotent response to be activated instead of the correct response.

**Structured event complex framework.** Most models of PFC function focus on the range of algorithms or processes that are mediated by the PFC, such as attention, working memory, conflict detection, etc. These processes are often discussed independently from the content upon which they act, which is considered to be stored in posterior and/or motor cortex. Grafman (2002; Forbes & Grafman, 2010) chooses instead to take a representational approach, viewing the PFC primarily as an organ of long term memory storage, and focusing on the structure and properties of the types of representations that are used by the PFC. This representational approach parallels the way in which posterior cortical functions, such as object, face or word recognition, are usually analyzed (Wood & Grafman, 2003).

Grafman's framework is based on the structured event complex (SEC), an ordered series of linked semantic representations of events. SECs have an onset, which is generally primed or activated by an environmental stimulus, as well as an offset, which corresponds to a behavioral goal or end state. The individual events that make up an SEC may be semantically independent, but an SEC is encoded and retrieved as a linked episode, and may be "run" as a simulation (cf. Barsalou et al, 2003).

SECs lend semantic and temporal structure to goal-directed actions and predictions, allowing an agent to project how various scenarios will unfold and make decisions on that basis. They may also be divided and combined in unique ways to allow for an increasing complexity of behaviors and predictions.

In Grafmans's view, SECs are employed throughout the PFC, with different areas specializing in SEC representations of different domains. In this way, posterior PFC SECs may represent simple, well learned action sequences while anterior areas deal with more complex SECs such as long term goals; medial PFC may utilize SECs that are predictive of behavioral goals while lateral SECs are composed of abstractions that may be adapted to a variety of applications, and so on.

**Guided Activation.** Miller and Cohen (2001) proposed an influential model that sought to integrate a wide range of data and a variety of previous frameworks of prefrontal function. In their view, the basic role of the PFC is to bias activation in other cortical areas toward behaviorally relevant goals. It accomplishes this by maintaining activity that is representative of the goals, rules and attentional templates involved in accomplishing a task. This activity provides context that guides the flow of activity elsewhere in the cortex along pathways that lead toward the accomplishment of those goals by establishing the appropriate mappings between inputs, internal states and outputs. Through repeated practice of a task, other brain structures such as premotor cortex learn these mappings, and so the need for prefrontal control diminishes; PFC activity is therefore only necessary when learning novel tasks.

The Stroop task provides a simple example of this. The correct response in this task is to name the color that a color name word is printed in; the prepotent, well-learned response, however, is to read the color name word itself. According to the guided activation model, prefrontal activity may bias activation toward the visual color feature representations in posterior cortex, and therefore provide those representations with greater influence over response activation, enabling the color naming response to outcompete the prepotent word reading response. The PFC provides top-down biases upon sensory representations, response execution, episodic memory retrieval, emotional valuation, etc., which may favor weaker, task-relevant responses or interpretations over others that are stronger, but task-irrelevant. This is especially important when task demands run counter to well learned responses, as in the Stroop task, and when they change rapidly, as in the WCST (and of course in many real life situations).

From the perspective of this model, active maintenance of representations is the primary distinguishing feature of prefrontal cortex. It is the underlying mechanism behind attentional set, working memory, and the inhibition of inappropriate responses, and allows for the integration of perceptions, actions, and consequences over time. This raises several questions. How is PFC activity maintained? How is a new representation 'gated' in, to begin being actively maintained? Does active maintenance of a representation end by gradual decay, or by being explicitly shut off or replaced? How does the prefrontal cortex determine from moment to moment what information should be gated in and maintained, and what information should not? How is more than one representation actively maintained at one time, and how may these multiple representations be updated independently of one another?

Miller and Cohen begin to address some of these questions. Several possible methods are presented by which the PFC could maintain the activity of a representation over time: cellular models assume that prefrontal neurons have an intrinsic mechanism for maintaining activity, while circuit models propose recurrent connectivity through closed loop attractor networks. Such loops may be fully contained within the PFC, or they may involve subcortical structures such as the basal ganglia and thalamus.

Based on previous research (Cohen et al, 1996; Braver & Cohen, 2000), the authors proposed that the midbrain dopaminergic (DA) system plays an important role in the updating or 'gating in' of new representations to be actively maintained. The level of activity of DA neurons appears to act as a reinforcement learning signal in the prefrontal cortex (Schultz et al, 1993; Schultz et al, 1997; Schultz, 1998; Schultz et al, 2000); DA neurons normally fire tonically, but produce a burst of activity when an unexpected predictor of reward occurs, or a temporary decrease in activity when an unexpected predictor of punishment (or an absense of expected reward) occurs. These signals initially accompany unexpected primary reward or punishment (such as feeding or pain), but through a learning process the signals 'migrate' backward in time to the initial cue that predicts the primary reinforcer (such as a sound or the presence of an object, which is predictably followed by the primary reinforcer). The moment that an unexpected reward predictor occurs is just when it would be most useful to gate that stimulus representation into active maintenance, so that it can bias attention and behavior in the direction of pursuing the predicted reward. Miller and Cohen suggest that the DA burst itself may act as the trigger to gate current input into active maintenance, by modulating the influence of afferent connections to the PFC.

The notion that the DA burst provides the PFC with a signal both for gating and for learning raises the possibility that it also allows the system to learn *when* to gate. Initial, exploratory, DA-triggered gating would sometimes produce behavior that leads to reward, and this would result in reinforcement of the association between the current context of cortical activity and the DA gating signal. This effect would increase the probability that a similar context in the future would again lead to gating, and then to rewarding behavior, further strengthening the association, and so on. This bootstrapping mechanism would allow the prefrontal cortex to learn on its own to adaptively control the gating of active maintenance. Such a self organizing control principle for the primary function of prefrontal cortex goes a long way toward closing the theoretical gap of the "central executive," and eliminating any need to invoke a mysterious "homunculus" at the top of the hierarchy of control.

While leaving many questions still to be answered, the guided activation model succeeds in integrating a large body of neurophysiological and neuropsychological data, and providing an elegant explanation for the many prefrontal functions addressed by earlier models. One important area to which this model has little new to add, however, is the study of prefrontal representations. The model is compatible with earlier views of PFC representations such as Norman and Shallice's schemas and Grafman's SECs, as discussed above. Miller and Cohen consider the PFC, like the rest of cortex, as being specialized for extracting the regularities across individual episodes, and generating abstract representations that may be applied in many different situations. Like Fuster, Grafman, and others, they acknowledge that PFC and motor representations require a method to enforce the correct sequencing of cognitive and motor actions. But the specifics of how abstract and time-spanning representations are generated by prefrontal cortex is not addressed by this model.

**Comparing Frontal and Posterior Cortex**

According to the guided activation model, and consistent with the other major frameworks reviewed above, one key function that sets frontal cortex apart from posterior cortex is the active maintenance of representations. This is carried out in such a way that it meets the seemingly contradictory goals of being robust to distraction, while also allowing representations to be dynamically gated into active maintenance as needed. A second important function of frontal cortex is the top-down gating of action selection through the executive hierarchy, from prefrontal through premotor and finally to motor cortex, so as to activate only the most appropriate task representation, preparatory set and specific action at the correct time.

There are, then, at least two important functions performed by frontal cortex that have no clear analogue in posterior cortex, both of which involve the gating of influence from one cortical area to another. Starting from a default assumption that there is a common cortical algorithm carried out by all areas of cortex, it would follow that there must be some variation or extension of this algorithm that would explain the functions specific to frontal cortex. In order to understand these functional differences, it would be useful to compare the structure and connectivity of frontal and posterior cortex.

The cytoarchitecture of prefrontal cortex is considered homotypical isocortex (Mesulam, 1997), having a structure very similar to that of posterior association areas. In areas of PFC that approach the bordering limbic regions, there is a gradual transition toward paralimbic features such as a thinner layer 4 and a decrease in the size of pyramidal cells in layers 3 and 5 (Barbas & Pandya, 1989). The motor cortex (area 4) is the only idiotypical area in the frontal lobe; it is missing a well developed granular layer (layer 4). Outside of these exceptions, the majority of frontal cortex is fairly homogenous in cytoarchitecture, and very similar to posterior areas that do not receive primary sensory afferent (Kaufer & Lewis, 1999). Given its largely uniform structure, functional specializations of the prefrontal areas are more likely to result from differences in connectivity rather than cytoarchitectural characteristics (Fuster, 1997).

As with posterior cortex, neurons in frontal cortex are organized in closely interconnected minicolumns that run perpendicular to the surface of the cortex. Neurons making up one minicolumn tend to be isoresponsive, tuned to fire in response to the same stimulus – for example, the same spatial location, visual feature or direction of movement – though the response timing relative to stimulus onset may vary between neurons within a minicolumn (Kritzer & Goldman-Rakic, 1995; Rao et al, 1999).

Frontal cortex also contains larger functional groupings, similar to the hypercolumns seen in posterior sensory areas. In frontal cortex these larger groupings are arranged in the form of elongated bands or stripes (Levitt et al, 1993). Some neurons located in frontal layer 2/3 project axon collaterals horizontally. These projections terminate in fields with the appearance of stripes, interleaved with gaps of similar size and shape. Cells within a single stripe or within a small number of nearby stripes may be interconnected in this way. These connections are reciprocal, and are formed between excitatory pyramidal cells. This recirculating arrangement within a group of stripes provides a possible anatomical substrate for the kind of reverberatory attractor network that would be useful for maintaining activity of a prefrontal representation. A possible complementary system of intrinsic maintenance of prefrontal activity is the property of bistability that has been found in prefrontal neurons, whereby a cell may be temporarily "switched" into a state that raises or lowers its threshold for activation (Frank et al, 2001). The NMDA receptors that may allow a prefrontal neuron to be switched into an "on" state, temporarily biased toward activation, are sensitive to the timing of incoming spikes (Wang, 1999). This raises the possibility that the temporal characteristics of afferent activity may determine whether or not that activity is "gated in" to influence active maintenance in the receiving stripe of prefrontal cortex. An actively maintained representation may remain stable under normal conditions of afferent activity, but may be replaced by a new pattern when subject to bursting afferent activity (O'Reilly, Munakata et al, 2012; Hazy et al, in preparation, as cited in O'Reilly, Hazy et al, 2012). This property will act as an important foundation for the models reviewed below.

Corticocortical connections in posterior sensory cortex are arranged in a clearly hierarchical manner. At each step in a sensory hierarchy, layer 2/3 cells project axons that connect with layer 4 cells at the next higher level of the hierarchy. Layer 5 cells, on the other hand, project back to the next lower level in the hierarchy, where their axons terminate in layer 1. Each step in such a hierarchy is reciprocally connected in this way. Frontal cortex is arranged similarly (Fuster, 1997). Primary motor cortex is at the bottom of the executive hierarchy. Here, layer 5 cells project axons top-down to the cranial nerves and the corticospinal tract, where they directly control voluntary movement. Being at the base of the motor hierarchy, primary motor cortex receives no bottom-up afferent and has no distinct layer 4. Primary motor cortex does project bottom-up from its layer 2/3 to layer 4 in premotor cortex, which reciprocally projects top-down from layer 5 to layer 1 in motor cortex. Likewise, premotor cortex projects bottom-up from layer 2/3 to layer 4 in adjacent areas of prefrontal cortex, which in turn project top-down to layer 1 in premotor cortex. However, this clear frontal lobe hierarchy ends once it reaches the prefrontal cortex; different areas of prefrontal cortex are not interconnected in a clearly hierarchical manner. The roles of individual layers remain the same in PFC – for example, layer 4 remains primarily a recipient of projections from other areas, and layers 2/3 and 5 project axons to other areas. But for two given nearby areas of PFC, the layer 4 of each area may receive afferent from the other area, and the layers 2/3 and 5 of each area may project to the other area (Barbas, 2006). The result is a highly interconnected, and sometimes circular, arrangement of pathways within prefrontal cortex.

Besides the hierarchically arranged connections between frontal regions and the heterarchical connections within the PFC, frontal cortex also has reciprocal cortico-cortical connectivity with posterior cortex. At each level of the posterior sensory hierarchies, bottom-up projections are sent both to the next higher level in the sensory stream, as well as to a specific area of frontal cortex (Fuster, 1997). For example, primary somatosensory cortex projects not only to secondary somatosensory cortex but also to primary motor cortex. Secondary somatosensory cortex (area 5) projects both to tertiary somatosensory cortex (area 7) as well as to premotor cortex. Area 7 projects to sensory convergence zones in the temporal lobe, as well as to dorsolateral prefrontal areas 45 and 46. Each of these connections are reciprocal. This arrangement results in multi-tiered connections between the posterior sensory hierarchy and the frontal executive hierarchy. Such nested feedback loops may facilitate the gradual migration of stimulus-response learning from prefrontal deliberate control to premotor automation, and may allow attentional set to be controlled at various points along the axis of specificity vs. abstraction.

In addition to cortico-cortical connections, the prefrontal cortex has many connections with subcortical structures, both directly and via the thalamus. The PFC has direct reciprocal connections with the brainstem, hypothalamus, amygdala, hippocampus, and paralymbic structures such as the cingulate cortex. The orbitofrontal PFC has extensive connections with the amygdala, in support of its "hot" emotion- and motivation-based processing, while the dorsal PFC, having a role in the encoding and retrieval of episodic memory, is directly connected with the hippocampus and parahippocampal cortex (Nauta, 1964; Fuster, 1997).

The PFC also receives its most prominent subcortical projections from the thalamus. While several thalamic nuclei project to the PFC, the majority of thalamic afferent is received from the mediodorsal nucleus (MD); in fact, one of the most widely used delineations of prefrontal cortex, both in application to humans and across species, is that area of cortex which receives afferent from mediodorsal thalamus (Fuster, 1997). The nucleus is made up of two components that vary in cytoarchitecture: the medial component is called magnocellular, due to the large size of its cells, while the lateral component with its smaller cells is called parvocellular. Prefrontal afferent channeled through the magnocellular component originates mainly from areas of the temporal lobe including  the amygdala, prepiriform cortex and inferior temporal cortex. Afferents channeled through the parvocellular component, on the other hand, originate from the prefrontal cortex itself. Two other thalamic nuclei, the ventral anterior nucleus (VA) and the ventral lateral nucleus (VL) are connected in a similar fashion with premotor and motor areas of the frontal lobe. These thalamic nuclei, which receive topologically organized excitatory projections from areas of frontal cortex and return excitatory projections to nearby areas of frontal cortex, comprise a potential candidate for the type of gating system that would be required by the models discussed above. If these thalamic relays could be opened or closed with behaviorally appropriate timing, the resulting mechanism could allow a representation active in prefrontal areas to be selectively gated into clusters of recurrently connected stripes, influencing and modifying the continuously maintained activity within those stripes by replacing it with a new representation. Likewise, such a gating mechanism could be used to selectively determine when the activity of one area of frontal cortex would be allowed to influence the activity of a second area, so as, for example, to allow the representation of a motor plan in premotor cortex to influence activity lower in the motor hierarchy, thereby initiating the execution of a planned behavior. In fact, there is increasing evidence that these thalamic relays may be used as a gating mechanism in those ways.

If such topologically mapped reciprocal connections between frontal cortex and thalamus do function as a gating mechanism, then the question remains as to how these gates would be controlled so as to execute behaviorally adaptive cognitive and motor actions. The answer may lie with a group of related subcortical structures called the basal ganglia (Redgrave et al, 1999; Frank et al, 2001; Sherman & Guillery, 2001). Nearly all areas of cortex project to the basal ganglia, by sending collateral axons from layer 5 pyramidal cells to either the caudate nucleus or the putamen (together called the striatum), which act as the input nuclei of the basal ganglia (Swanson, 2000). Unlike most cortical efferents, projections to the striatum are not reciprocated; the basal ganglia do not project back to cortex directly. Instead, the basal ganglia project inhibitory efferent to the very thalamic nuclei that mediate reciprocating loops from frontal cortical areas and back again (MD, VA and VL), as described above. Normally the basal ganglia send a tonic inhibitory signal to these thalamic nuclei, but by pausing this inhibitory influence or by increasing its intensity, the basal ganglia is able to modulate the degree to which cortical input is "passed through" and relayed back to cortex (Chevalier & Deniau, 1995). In this way, the basal ganglia system is in a position to control the thalamic gating mechanism of the frontal cortex.

This basal ganglia control mechanism is not a single, unified circuit; instead, it is comprised of at least five distinct cortico-striato-thalamo-cortical loops that operate in parallel (Alexander et al, 1986). In the case of each of these loops, a discrete area of the striatum receives topologically mapped afferent from a particular area of frontal cortex, as well as from several functionally related areas of frontal or posterior cortex. This area of the striatum then sends inhibitory projections to specific areas of the globus pallidus, which in turn send inhibitory efferent to the area of thalamus responsible for relaying information to the particular area of frontal cortex that makes up the primary input for this particular loop (Figure 3).

The most well researched of these control loops is the motor circuit. The supplementary motor area, which is known to play an important role in the programming and initiation of movements, projects efferent from its layer 5 pyramidal cells to the putamen (a section of the striatum). Other cortical areas that represent information relevant to motor control, including the arcuate premotor area, motor cortex, and parietal somatosensory areas, also project to the same region of the striatum. The putamen projects to the globus pallidus, which in turn sends inhibitory projections to the ventral lateral nucleus of the thalamus, which relays motor information to the primary and supplementary motor areas (Alexander et al, 1986; Strick, 1976). In this way, the basal ganglia is able to utilize information from motor and somatosensory areas to inform the control of the thalamic nucleus that gates motor information into motor and supplementary motor cortex.

In addition to the motor circuit, an oculomotor circuit has been identified which receives information from the frontal eye field, the dorsolateral PFC and the posterior parietal cortex, and controls the thalamic nucleus (the superior colliculus) that relays back to the frontal eye field; a prefrontal circuit which receives information from the dorsolateral PFC, posterior parietal, and promotor areas, and controls the thalamic relay back to the dorsolateral PFC; an orbitofrontal circuit which receives information from the orbitofrontal cortex, temporal cortex and anterior cingulate cortex (ACC) and which controls the thalamic relay back to orbitofrontal cortex; and an anterior cingulate circuit that receives information from the ACC, hippocampus and entorhinal cortex and controls the relay of information back to the ACC. Somatotopic arrangement is maintained through all stages of these circuits; for example, areas of somatosensory, premotor and motor cortex that map to the left hand all project to the same section of striatum. This section of striatum, via further connections within the basal ganglia, projects to the area of VL that relays information to motor areas mapping to the left hand. The spatial resolution of this gating control mechanism remains unknown; while there are at least five distinct circuits, it's possible that independent gating actions may be performed at a much finer scale, possibly gating thalamic input to individual clusters of stripes.

In summary, the primary functional differences between frontal and posterior cortex can be viewed as comprising a frontal mechanism of recurrent connections and bistability that supports the active maintenance of representations, and a system of thalamic relays that allow a flow of information to be dynamically gated into active maintenance, as well as from one area of frontal cortex to another. The basal ganglia adaptively controls this gating system, learning how to respond to the diverse cortical afferents that it receives by opening and closing thalamic relays at behaviorally appropriate times. How the basal ganglia make use of the brain's DA system to control the gating process that underlies cognitive and motor executive function is one of the issues addressed by the neural models reviewed below.

**Neural Models of Frontal Cortical Function**

Based on the ideas described above, a neural model was proposed by Frank et al (2001) and further extended (O'Reilly & Frank, 2006) as a computational model, the prefrontal cortex and basal ganglia working memory model (PBWM). These models elaborate and simulate several neural systems underlying frontal cortical function, and have been successfully employed in exploring both normal and impaired frontal function by comparing performance and network dynamics on frontal tasks between the model and human subjects (e.g., Frank et al, 2004; Frank & O'Reilly, 2006; Frank et al, 2007; Moustafa et al, 2008; Frank & Badre, 2012).

A summary of the proposed neural model is as follows. The striatum, as described above, receives input from a wide range of cortical areas; different sections of the striatum receive input from different groups of functionally related cortical areas, depending upon which thalamic relay that section of the striatum is involved in controlling. The striatum itself is composed of an interleaved mix of two different types of cell clusters, called patch (or striosomes) and matrix (or matrisomes). Striosomes are involved in controlling the brain's DA system, and will be discussed further below. Matrisomes contain two types of spiny neuron, referred to as "Go" and "NoGo" for their functional roles in the gating system. Go neurons send inhibitory projections to the internal segment of the globus pallidus (GPi), which in turn sends inhibitory projections to the corresponding thalamic relay area. Therefore, when the striatal Go cells are activated, they inhibit the connected GPi cells (which are normally tonically active) and so remove inhibition from the corresponding thalamic relay cells. This allows those thalamic relay cells to activate based on the cortical input that they receive, and so opens the thalamic "gate" that those cells represent functionally. This series of two inhibitory steps that results in the opening of a thalamic gate, the direct "Go" pathway, is complemented by an overlapping series of three inhibitory steps that result in the closing of a gate. In this indirect "NoGo" pathway, the NoGo matrisome neurons send inhibitory projections to the external segment of the globus pallidus (GPe), which itself normally sends a tonic inhibitory signal to the corresponding Go pathway section of the GPi. Therefore, when the striatal NoGo cells are activated, they quiet the tonic activity of the corresponding GPe cells, which thus stop sending an inhibitory signal to the corresponding GPi cells. The GPi cells are then free to increase the strength of the tonic inhibitory signal that they send to the corresponding thalamic relay cells, preventing activation of those cells and so closing the thalamocortical gate that they represent.

This system of opposed Go and NoGo pathways places the locus of control for frontal thalamic relays with the striatal matrix cells. The activity of striatal matrix Go cells will bias the corresponding thalamic gate toward opening, while the activity of striatal matrix NoGo cells will bias the corresponding gate toward closing. What remains is the question of how these striatal matrix cells are themselves controlled. As discussed above, the midbrain DA signal that responds with increased firing to the unexpected presence of a positive reinforcer would be a good fit for the requirements of a working memory gating signal; it is just those stimuli that result in the unexpected prospect of reward that should be maintained in memory so as to bias behavior toward the pursuit of that reward. In addition, it would be useful for the gating mechanism to learn what contexts of cortical activity, when followed by the opening or closing of a thalamocortical gate, result in unexpected reward – so that gating regimens can be learned and can continue to be followed even once the rewards involved are no longer unexpected, and thus the DA system is no longer engaged.

The striatal matrix cells are well equipped to fulfill both of these requirements. The substantia nigra pars compacta (SNc), another component of the basal ganglia, continuously releases DA into the striatum, at a level that varies with the unexpected presence or absence of reinforcers. Striatal matrix cells incorporate two different types of DA receptors: Go cells have D1 receptors, making them more likely to fire in response to cortical afferent in the presence of DA, while NoGo cells have D2 receptors, making them less likely to fire in the presence of DA. Thus, the higher DA concentrations that signal the presence of an unexpected positive reinforcer will bias the striatal Go cells toward activating (given that they are also receiving cortical afferent) but will simultaneously bias the NoGo cells against activation. The DA signal is therefore able to directly control the opening and closing of thalamocortical gates. In addition, because DA also modulates the activity-dependant plasticity of synapses, at the same time that the DA signal is directly controlling gating, it is also teaching the matrix cells how to control gating on their own in similar future circumstances, by strengthening the synaptic associations between the cortical input to the striatum (representing internal and external context that may be relevant to determining an adaptive gating strategy) and the Go or NoGo matrix cells that are activated as a result of the current DA signal. Therefore the DA signal not only controls gating directly, but also teaches the striatum how to control gating in similar future circumstances when the DA system may not be engaged.

Up to this point we have traced the control of the thalamic gating functions that mediate action selection and working memory updating back to the midbrain DA system. This system is commonly considered to signal reward prediction error (Rescorla & Wagner, 1972). Midbrain DA cells (such as those of the SNc that project to the striatum) normally fire tonically, but the firing level transiently increases when an unexpected primary positive reinforcer occurs (e.g., a squirt of juice is often used in experiments with monkeys). If such a positive reinforcer is consistently preceded by another stimulus (e.g., a flashing light) then the DA burst will gradually "migrate" from the primary reinforcer to this predictive stimulus, which has become a secondary reinforcer. Once training is complete, the DA burst will occur only when the secondary reinforcer is presented; DA levels will remain constant when the primary reinforcer follows it, because the primary reinforcer is now fully expected. Should the secondary reinforcer then be presented but *not* followed by the primary reinforcer, there will be a dip in DA level at the time when the primary reinforcer is expected, indicating negative reward predicton error; the primary reinforer was expected but did not occur. The system works inversely for negative reinforcers; there is a dip in DA level when an unexpected negative primary or secondary reinforcer occurs, or a DA burst when an expected negative reinforcer does not occur.

The dominant model of this system of signaling reward prediction error is the temporal difference (TD) learning algorithm (Sutton, 1988; Sutton & Barto, 1998). The TD algorithm determines the reward prediction error by taking the difference between the actual reward value at the current time plus estimated future reward (discounted in proportion to its distance in the future), and the estimated reward value for the current time. The reward prediction error is then used to modify the weights that determined the reward estimate for the current time, so that in a similar future circumstance they will produce an estimate that more closely resembles the current actual reward value. By repeating this process over multiple trials, a reward prediction (corresponding to a DA burst in the brain) will incrementally step backward in time to correspond with the earliest secondary reinforcer that predicts a primary reinforcer.

Although the TD learning algorithm was developed before DA response was measured in the brain, it closely reflects the dynamics of the midbrain's DA system. Because of this, early versions of the PBWM and associated DA gating models (e.g., O'Reilly & Munakata, 2000) leveraged the TD algorithm to generate the DA signal used by the model for learning and gating. However, the TD algorithm has several shortcomings. Because it relies on the sequential chaining of predictions through earlier and earlier time steps, any unpredictable event inserted within the learned sequence of events can break the chain and prevent learned reward predictions from being applied in the new situation. Biological reward prediction, on the other hand, is robust to the insertion of unpredictable events. The sequential chaining of TD learning also predicts that DA response would incrementally shift backward in time as a secondary reinforcer is learned, "travelling" backward over the intervening gap between the primary reinforcer and the secondary reinforcer. Instead, the biological DA response shows a gradual decrease in DA response to the primary reinforcer accompanied by a gradual increase in response to the secondary reinforcer, without any interpolation of the response through the intervening time. Finally, the TD learning algorithm does not have any clear functional mapping to biological components.

For these reasons a biologically grounded model of DA response was developed, the primary value and learned value Pavlovian learning algorithm, or PVLV (O'Reilly et al, 2007). This model proposes that DA response is based on an opponent process involving two distinct brain systems that drive DA release in response to primary and secondary reinforcers, respectively, and a third system that inhibits DA release in response to any reinforcer that is already expected. The lateral hypothalamus is hypothesized to signal midbrain neurons in the SNc and ventral tegmental area (VTA) to release DA in the areas to which they project, in response to the presence of a primary reinforcer. The central nucleus of the amygdala is thought to perform a similar role for secondary reinforcers, by learning what cortical representations are reliably associated with primary reinforcers and signaling the SNc and VTA when such representations are active. These systems are complemented by cells in the striosomes of the ventral striatum which learn what cortical representations reliably predict both primary and secondary reinforcers, and, via projections to the SNc and VTA, inhibit DA release at the time when a reinforcer is predicted to occur. This results in a DA burst occurring only in the presence of a primary or secondary reinforcer that has not been predicted, because any potential DA burst is cancelled out if the reinforcer is predicted. This also results in a dip in the tonic DA release when a reinforcer is predicted, but does not occur. Learning in this system occurs by enabling Hebbian modification of synaptic weights only when a primary reinforcer is present or expected. The PVLV model is able to bootstrap a DA response system that mimics the dynamics of the biological system more closely than does the TD learning algorithm, and eliminates the TD algorithm's dependence on a fragile learned chain of events. In addition to training itself in this way, the DA response system generates the signal that is used by the basal ganglia to control and train the frontal thalamocortical gating system, driving cognitive and motor actions in a behaviorally adaptive manner that maximizes positive reinforcement and minimizes negative reinforcement.

The neural models underlying PBWM and PVLV were abstracted and implemented as computational models using the Leabra point neuron connectionist architecture (O'Reilly & Munakata, 2000; O'Reilly, Munakata et al 2012; O'Reilly, Hazy et al, 2012), an evolution of the influential parallel distributed processing (PDP) framework (Rumelhart et al, 1986; McClelland et al, 1986). A Leabra network typically employs multiple regions of model units connected bidirectionally and hierarchically.

Within a region, local inhibition (simulated using a kWTA function) enforces sparse activation; this competitive effect combined with Hebbian learning results in sparse distributed representations in which individual units come to represent distinct high probability features or combinations of features (e.g., Rumehart & Zipser, 1985). This aspect of the architecture closely resembles the CLA's spatial pooler, as described above. An individual Leabra unit is considered to correspond in its connectivity and response properties to a single cortical minicolumn (O'Reilly, Hazy et al, 2012); this differs from the CLA, which treats the different cells within a minicolumn as representing the same feedforward input but within different temporal contexts. Whereas the CLA's temporal pooler employs this system to generate invariant representations in an unsupervised manner by superimposing the highest probability predictions that result from a given input, Leabra models rely on the addition of error-driven supervised learning to generate invariant representations. However, the issue of how invariant representations are generated may be put aside for the purpose of this discussion, as the basic PBWM and PVLV models are trained entirely using Hebbian synaptic modification, modulated by a DA signal in order to effect reinforcement based learning.

An instructive example of the function of the PBWM and PVLV models is in their application to a response task with hierarchical structure (Frank & Badre, 2012). The task, adapted from an earlier study performed with human subjects (Badre et al, 2010), involves a series of trials in which the subject is shown an image, and then, after a short delay, must press one of 3 buttons in response (Figure 4). Every image is selected from the same set of 18 images, each of which maps uniquely to one correct response. After each trial the subject receives feedback on whether their response was correct, with the goal of learning the mappings and providing as many correct responses as possible. Images vary along three dimensions: there are three object shapes, three object orientations, and two possible colors of a square that frames the object. The experiment consists of two epochs. In the first epoch, called the "flat" condition, there is no structured rule that governs the mappings of image to correct response, and so each mapping must be learned individually. In the second, "hierarchical" condition, however, there is hierarchical structure to the mapping rule, which may be exploited by the subject in order to speed up learning of the mappings: if an image is framed in one color, then its object shape alone determines the correct response mapping; if it is framed in the other color then its object orientation alone determines the mapping (Figure 5). If this hierarchical structure is exploited, then fewer individual mappings need be learned. A model based on PBWM and PVLV was constructed for this task, and both its performance and neural dynamics were compared with those of human subjects.

As an approximation of the thalamic relay gating system, a simplified basal ganglia model is used which is composed of a set of Go and NoGo striatal units corresponding to each representation being gated. The relative activation of Go versus NoGo units determines whether a gate is opened or closed. The activity of each Go or NoGo unit is driven by input it receives from one of the model's cortical regions. The synaptic weights of these input connections are trained using a Hebbian rule modulated by the DA signal generated by the PVLV model, which is based on the positive or negative feedback received after each trial. DA bursts bias Go cells toward activating, and train their input connections so that they will be more likely to activate when they receive similar cortical input in the future. Conversely, DA dips bias NoGo cells to activate, and train their input connections. The result is that the basal ganglia model learns to open and close its gates in response to patterns of cortical input, so as to maximize positive reinforcement. Each gate system used in this model has three separate sets of striatal units, so that it can independently gate output to three different sets of cortical units (corresponding to separate "stripes" of frontal cortex) used to represent the three visual attributes of a stimulus image (object shape, object orientation, and frame color). This allows a gate system to pass one or more attributes through to influence response selection in a given trial, while not passing through other attributes that are irrelevant to response selection and could otherwise interfere if passed through.

The architecture of this model's network is shown in Figure 6. For this task, four separate gate systems are employed, to perform several different functions. The model includes a visual input region with eight localist units, one representing each possible attribute of a stimulus image (three object shapes, three object orientations, and two frame colors). There is also an output region, with three localist units representing the three possible responses. These regions are connected via the first thalamic gate, which is used to control response selection based on the probabilities of positive reinforcement given a representation of the current input state which it receives from another region that corresponds to the output of dorsal premotor cortex. In this model, dorsal premotor cortex is represented as two separate regions. The first, PMd maintain, corresponds to layers 2/3 of premotor cortex and is capable of active maintenance of the input it receives. It receives afferent from the model's visual input area, and this afferent is input gated by a second thalamic gate. This gate learns which stimulus attributes should be gated into active maintenance within PMd maintain, based on the probability that their maintenance will lead to positive reinforcement. A second region, PMd output, corresponds to layer 5 of the same area of premotor cortex. The flow of information from PMd maintain to PMd output is gated by a third thalamic gate. In this way, the updating of an actively maintained representation is separated from the output of that information to other areas. It is then PMd output that provides contextual information to the response gate that controls action selection.

In order to be able to exploit the response rule structure of the hierarchical task condition, the model includes another active maintenance region representing a more anterior section of premotor cortex, and capable of storing higher level task information that is used to contextualize gating at the lower level. This prePMd output region receives input directly from the model's visual input region, but that input is gated by a fourth thalamic gate. This gate learns what stimuli, when maintained in this higher order context buffer, is predictive of reward. The output of prePMd output is then sent to the PMd output gate, to provide context for its decisions about what stimulus dimensions to expose to the response gate.

When there is latent hierarchical structure in the response rules, the architecture of this model allows it to discover and exploit that structure. The prePMd gate learns to gate the frame color information into active maintenance in prePMd output. The frame color determines whether it is the object shape or orientation that maps to the correct response, and so the frame color is high level task information that is useful for contextualizing the lower level decision of which stimulus attribute to base a response upon. Because the maintenance of this information in prePMd output tends to lead to a greater probability of correct response and of positive reinforcement, the prePMd gate learns to gate in and maintain just the frame color information. In the same way, the PMd input gate would learn to maintain only the object shape and orientation information, and the PMd output gate, responding to the frame color information sent to it by prePMd output, would learn to output only the task appropriate object attribute: object shape in the case of the first frame color, or object orientation in the case of the second frame color. The response gate will now receive from PMd output only the information about the relevant object attribute, which greatly simplifies the task of learning the correct response mappings.

This computational model embodies a number of the functional principles common to the frameworks of frontal cortical function discussed above. Its abstracted thalamocortical gates are employed in several different ways that correspond to their proposed roles in modulating activity in the frontal cortex. The response gate is used to select a motor action and activate the representation of that action in the motor area. The PMd input gate and the prePMD gate are each used to select which representation will be gated into active maintenance within a cortical area. The PMd output gate selects which maintained representations will be allowed to influence other brain areas downstream, in the manner of a top-down attentional effect, and enables representations to be actively maintained without necessarily being output to other areas. Finally, the separate but complementary functions of the PMd and prePMd regions, such that the prePMd region maintains high order task information while PMd maintains lower order task information and uses the high order information to determine what lower order information to project downstream, reflects empirical findings that more anterior areas of frontal cortex are involved in learning abstract, higher order task information (e.g., Koechlin et al. 2000, 2003; Christoff et al, 2009). Learning in each gate at every level of this model is modulated by the DA system's reward prediction error signal, such that DA bursts promote Go learning and DA dips promote NoGo learning. An initial bias toward Go activity promotes exploratory gating activity, and launches a "walk" through the space of gating responses in search of maximized reward. This bootstrapping mechanism is effective in guiding executive control of working memory, attention and response toward exploiting the hierarchical rule structure in this task.

In summary, while many aspects of frontal cortical function remain poorly understood, a prevailing perspective holds that frontal areas differ from one another principally in their connectivity, which determines the nature of the representations that they are capable of generating, as well as maintaining in activation, so as to influence processing in connected regions. The recurrent thalamocortical relays that are unique to frontal areas, and that appear to be modulated by the basal ganglia reinforcement learning system, provide a gating mechanism, similar to a production system, by which the flow of information between frontal areas may be adaptively switched on or off with high spatial resolution. Converging lines of evidence suggest that this context sensitive gating system mediates competitive action selection, the updating of working memory, and the control of top-down attentional effects. While several other models have been developed based on a similar biological and computational framework (e.g., Brown et al, 2004;

Stewart et al, 2012), the family of models to which PBWM belongs is a particularly well developed and diversely applied embodiment of this set of principles.

## Toward a Comprehensive Model of Cortical Function

Though it plays a primary role in those aspects of perception, cognition, and behavior that are most strongly identified with human intelligence, the cerebral cortex is a fairly recent evolutionary addendum to a labyrinthine organization of subcortical structures. As such, it is not unexpected that the cortex, far from being an independent "ivory tower" of higher intelligence, works in close collaboration with its phylogenetically more ancient neighbors. Both posterior and frontal cortex engage with the hippocampus for the storage and retrieval of episodic memory, and with the cerebellum for the development of coordinated motor control. Posterior cortex receives preprocessed sensory information from the thalamus. Frontal cortex, being in charge of behavior, relies heavily on its interactions with subcortical nuclei such as the hypothalamus, amygdala and other limbic structures to inform the action selection process with motivation and emotion. In addition, the system of thalamic relay loops that is unique to the frontal cortex allows the flow of activity between areas to be selectively gated by the basal ganglia reinforcement learning system, so as to base both motor and cognitive actions in the adaptive pursuit of maximized reward and minimized punishment. Clearly, the function of the cerebral cortex cannot be comprehensively studied in isolation; it is an important part of a much larger system.

By understanding how cortical function depends upon interactions with these subcortical systems, we can approach the problem of separating out what specific processes a general cortical algorithm would be responsible for, if such an algorithm exists. As we have seen, the frontal cortex performs roles that are very different from those of the posterior cortex. These differences can substantially be explained, however, by differences in connectivity. Most crucially, the thalamocortical gating system mediates the unique role of the frontal cortex in driving cognitive and motor behavior. One area where there is strong evidence of an intrinsic functional difference between frontal and posterior cortex is with regards to the bistability and highly recurrent local connectivity that appear to underlie active maintenance of frontal representations (Levitt et al, 1993; Kritzer & Goldman-Rakic, 1995; Wang, 1999). In other respects, however, frontal and posterior cortex may simply be performing the same kinds of operations upon different inputs.

In this section I return to the consideration of what processes make up a plausible candidate for a general cortical algorithm. Using the HTM CLA as a starting point, I will examine how it may be adapted to fulfill the requirements of both posterior and frontal cortical function.

**Sequence Storage**

The basal ganglia have often been implicated in the storage of motor sequences (e.g., Graybiel, 1995; Cromwell & Berridge, 1996). However, more recent experimental evidence suggests that while the basal ganglia is involved in the learning of novel motor sequences, output from the basal ganglia is not required in order to carry out well learned motor sequences (Desmurget & Turner, 2010). In addition, conditions that compromise basal ganglia function in humans, such as Parkinson's disease, impair the ability to initiate action (as would be expected from the gating model discussed above) but do not impair the ability to carry out a motor sequence once action has been induced (O'Reilly & Munakata, 2000).

If motor sequences are not stored in the basal ganglia, then like perceptual sequences they are likely stored in the cortex itself, perhaps in cooperation with re-entrant thalamocortical loops (e.g., Granger, 2006). According to the CLA model, one of the main functions of cortex is to record the temporal context in which spatial patterns of activation occur, so as to be able to predict and reactivate those patterns at the appropriate time based on learned sequences of activity. Top-down connections from layer 5 of one cortical area to layer 1 of a hierarchically lower area may be organized in such a way that when a static pattern representing a temporal sequence is strongly activated in the higher area, each element of that sequence is played out over time in the lower area. In this way, output gating of a particular cognitive or motor sequence representation in an area of frontal cortex could initiate the execution, in hierarchically lower areas, of the learned sequence of actions represented by that single, static pattern of activity. This ability to collapse a sequence of events that is extended over time into a static pattern of activation at a higher level may lie behind one of the primary functions of the frontal cortex, that of integrating temporally separated events and contingencies (Fuster, 1997).

In addition, such a method of recording sequences and generating predictions based on temporal context may make an important contribution to the DA signal system discussed above. The PVLV model is based upon simultaneous associations rather than predictions (O'Reilly et al, 2007). Because of this, a stimulus representation must be actively maintained through to the time at which the primary reinforcement event occurs, in order for that stimulus to be learned as a secondary reinforcer. The frontal cortex active maintenance system is suggested to be responsible for this. However, it is not yet empirically clear whether in the brain a stimulus representation must be actively maintained in order to be learned as a secondary reinforcer. A CLA-like prediction mechanism, either implemented in the cortex or the amygdala, would allow predictive associations to be learned between a secondary reinforcer and a primary reinforcer, without the need for the representation of the secondary reinforcer to remain active until the primary reinforcer occurs in order for this learning process to take place.

**Corticocortical Connectivity**

While the Zeta 1 implementation of HTM incorporated a top down pathway of excitation in order to meet the requirements of its equivalence with Bayesian belief propagation, the CLA implementation has not yet been used with top down projections in any published study. Clearly in order to model a variety of cortical functions, including attentional bias and Bayesian inference, top down excitation would need to be integrated with the CLA.

Because CLA cells have three different states (inactive, active and predictive) the question of how to connect two CLA regions isn't as straightforward as it can be for networks with cells that are either active or inactive. Bottom up connections, corresponding to projections from layer 2/3 in one cortical area to layer 4 in a hierarchically higher area, would transmit excitation from both active and predictive cells. This is necessary in order for both the active input representation and the high probability predictions that result from that input representation to be passed forward as a single, overlaid whole. The feedforward projection of a single representation combining both the current input and its high probability predictions is at the heart of the CLA's abilities to generate invariant representations and to compress temporal sequences into static spatial patterns.

Top down connections, corresponding to projections from layer 5 in one cortical area to layer 1 (and from there to deeper layers) in a hierarchically lower area, have different requirements. While the feedforward pathway compresses time, the feedback pathway expands it. Top down connections should therefore transmit excitation from cells in the active state, but not from cells in the predictive state. In this way, as a temporal sequence plays out in a given cortical area, only the currently active representation in that sequence will be projected top down to a lower area. That lower area may itself represent a subsequence, and so on; multilevel sequences can be played out where each successively lower cortical level represents briefer but more detailed sequences than the level above it.

Many details of what would make up the most effective architecture of CLA regions in order to accurately model cortex remain unresolved. The several distinct cortical layers differ in connectivity and cytoarchitecture, and may represent various transformations of the same pattern. The hierarchical frontal model discussed above (Frank & Badre, 2012), for example, treats layer 5 as representing a copy of layer 2/3, but modulated by a thalamocortical output gating function. There is also the question of where and how the activation of successive elements of a sequence are timed; Hawkins et al (2010) propose that layer 5 uses a thalamic timing signal available in layer 1 to control the precise timing of outputs. Finally, feedforward projections from a single cortical minicolumn may be sensitive to one or a small set of individual cells, and so transmit representations that vary with the temporal context of the active efferent pattern, or projections may act disjunctively for a large group of cells within one column, and so transmit a representation that is invariant to the temporal context of the active efferent pattern. Each of these methods have their own advantages, and it may be that different cortical areas vary in this parameter.

**Synaptic Learning**

As described above, the CLA's spatial pooler performs competitive Hebbian learning, with the result being that individual columns come to specialize in responding to particular spatial patterns of input that occur with high probability. To ensure that most columns do learn to represent some input pattern (and so are not "wasted"), a boosting mechanism increases the input sensitivity of columns that have been active with very low frequency. Conversely, the input sensitivity of columns that have been active with very high frequency is attenuated, to prevent columns from overgeneralizing to represent too wide a variety of input patterns. The principle motivating this dynamic modulation of input sensitivity is well founded; it does improve the quality of learning, and a similar approach is taken in analogous learning algorithms such as vector quantization. However, the CLA implementation of this boosting mechanism (Hawkins et al, 2010) is poorly specified and lacking in biological grounding, which resulted in a variety of ad hoc approaches being used by different CLA implementations.

The Leabra framework (O'Reilly & Munakata, 2000; O'Reilly, Munakata et al, 2012), which incorporates competitive Hebbian learning very similar to that of the CLA's spatial pooler, is a mature biologically based approach to modeling brain systems that has been very successful in modeling a variety of phenomena including the frontal and subcortical reinforcement learning mechanisms discussed above. As such, it may offer insights that are transferrable to the CLA, particularly in areas of close overlap such as competitive Hebbian learning.

Recent versions of Leabra use a synaptic weight modification rule derived from a biologically detailed model of spike timing dependant plasticity (Urakubo et al, 2008). In Leabra's XCAL model, learning dynamics are captured by a piecewise linear function that is essentially a linearized version of the BCM synaptic learning model (Bienenstock et al, 1982). Synaptic weight change is a function of the product of the short term average activity of the presynaptic and postsynaptic neurons. If this value is above a given threshold, the weight of the synapse increases; otherwise it decreases. What is distinctive about this model is that it employs a floating threshold, whose value is based on the long term average activity of the postsynaptic neuron: the greater the long term activity, the higher the threshold, making weight decrease more likely than weight increase. Conversely, the lesser the long term activity, the lower the threshold, which favors weight increase over weight decrease. The result is a homeostatic drive toward a roughly equal distribution of activity among units, and so also a roughly equal division of the representational space. This achieves the same aim as the CLA's boosting mechanism in a principled manner that is based on detailed biological theory and parallels the proven BCM model. It would therefore be worthwhile exploring the possibility of incorporating this technique into the CLA's spatial pooler.

It is worth mentioning another point of connection between Leabra and the CLA. From the start, Leabra has utilized a combination of competitive Hebbian "model learning" and error driven "task learning". Competitive learning is well suited to the unsupervised extraction of statistically common patterns from input data, while supervised error driven learning is useful for teaching a network arbitrary mappings from input data to an output response, as in category learning. Earlier versions of Leabra used the GeneRec algorithm (O'Reilly, 1996) for error driven learning, a biologically plausible equivalent to the influential backpropagation algorithm (Rumelhart et al, 1986). Synaptic learning with GeneRec is fundamentally different from Hebbian learning. It requires that learning trials be composed of two phases. In the "minus" phase, the network is given an input and allowed to settle to its own resting activity state. In the "plus" phase, the supervised, "correct" output activity is clamped in an output region. Its influence feeds back through the network, which is allowed to settle to a new resting activity state given the additional output constraint. Changes in synaptic weights are then derived, for each trial, from the differences in the activity of units between the minus and plus phases.

Recent versions of Leabra have employed a much more elegant and biologically motivated way of integrating error driven learning with competitive Hebbian learning. It begins by reframing the role of error driven learning: more specific than learning input-output mappings, error driven learning may be thought of as generating predictions and then learning from the difference between those predictions and the actual outcomes (McClelland, 1994). By viewing error driven learning as a sequence of temporal events in this way, the same XCAL learning rule that results in Hebbian synaptic weight changes can also be adapted to error driven learning, by simply basing the threshold dynamics on a faster timescale. Rather than the threshold, which determines what level of activity is required to increase vs. decrease synaptic weight, being based on the long term average activity of the postsynaptic unit (which provides the homeostatic "boosting" effect for Hebbian learning), the threshold may be based on a medium term timescale and so reflect the predictive activity that occurs after an input is received but before the outcome is received. If a representation is predicted, its synaptic thresholds rise, making decreases in synaptic weight likely if the outcome does not match the prediction. If a representation is not predicted, its synaptic thresholds lower, making increases in synaptic weight likely if the outcome matches this unpredicted representation. Modifying the thresholds in this way results in error driven learning, such that future predictive activity in response to the same input will more closely reflect the actual outcome.

This error driven method of shaping the learning of predictions by contrasting predictions with outcomes is closely related to the process used by the CLA's temporal pooler. The Leabra system continues to depend on the use of separate populations of units to represent input and output, respectively. It no longer requires a strict separation between a minus and plus phase, but it does require some kind of learning signal to be provided when an outcome occurs, so that it may be contrasted with the immediately previous predictive activity and change synaptic weights accordingly. The CLA overcomes both of these limitations by proposing two separate cell states, active and predictive. This allows the same population of units to represent both prediction and outcome simultaneously, and so also to learn continuously rather than in discrete phases or when an explicit learning signal is provided. Nonetheless, the Leabra method of learning predictions using fast manipulation of synaptic learning thresholds may offer valuable insights and biological grounding to research using the CLA. It may be a hopeful sign of progress being made in the understanding of cortical function, that these two biologically based architectures appear to be moving in a direction of convergence.

**Abstract Representations**

One aspect of frontal cortical function that has been given less attention than it is due, both in this review and in the literature in general, is the biological basis of the generation and manipulation of abstract frontal representations. Response properties of some PFC cells have been shown to correspond with abstract rules (Wallis et al, 2001), and models such as those of Norman & Shallice (1980, 1986; Shallice, 1982; Shallice & Burgess, 1991, 1996), Grafman (2002) and Fuster (1997) have emphasized the requirement that frontal cortex dynamically generate and manipulate abstract representations of concepts, rules and actions, and use these representations to structure behavior. Computational models have been developed with the aim of combining the representational strengths of symbolic cognitive models with the dynamics, symbol grounding and biological plausibility of connectionist models (e.g., Jilk et al, 2008). However, there remains a wide gap in understanding separating the two perspectives. How are abstract representations of relationship roles, standing in not for perceptual objects but for a set of contextual relationships between objects (or even between other abstract roles), generated? How are other appropriate representations mapped to these roles, in the manner of variables filling slots (Anderson, 1983)? How are multiple representations, or even multiple instances of the same representation, combined compositionally into a common schema (Jackendoff, 2002)? The answers to these questions are still far from clear, but the possibility that the cortex implements an algorithm such as the CLA that employs temporal slowness to generate invariant representations suggests a novel approach to these issues.

As described above, the CLA generates low level visual representations with spatial invariance by overlaying the sparse active representation of a given visual feature with the predictive representations of those visual features that are subsequently active with the highest probability. For a low level visual feature, this would typically result in an overlay of the representations of the same visual feature at a number of nearby spatial locations (Figure 2). When the same process is applied to higher level representations of visual objects, each composed of groups of spatially invariant features, then because spatial invariance has already been extracted, the transitions of activity that would occur with the highest probability would be from one visual object to another nearby object, as the result of shifts in visual attention. In this way, the representation of a given visual object would come to include the predictive activity of representations of nearby visual objects. The visual representation of 'nose', for example, may come to include predictive activity of nearby objects such as 'eye' and 'mouth' that are commonly the subject of shifts in visual attention from 'nose'. Some fraction of the active cells that participate in the representation of a visual object would in fact be representing the visual context of the object, rather than the attributes of the object itself. Once this is the case, then so long as the contextual relationships are preserved, the visual attributes can be altered while maintaining a substantial part of the object's active representation. This would allow for the processing of visual metaphors such as the face of a snowman, where the individual visual attributes are changed but their contextual relationships with one another remain.

This same principle may be applied to frontal representations. In frontal cortex, transitions in activity from one representation to another do not necessarily depend on shifts in visual attention, but may also be controlled by the gating of thalamocortical relays as discussed above. Rather than transitioning with high probability between nearby objects, it is likely that frontal activity transitions with high probability between objects that are conceptually related due to being linked within the same frontal sequences or schemas. If a CLA-like mechanism is operating in frontal cortex, then the frontal representation of an object may come to include cell activity that is representative of that object's conceptual contexts, rather than just the attributes of the object itself. Different objects that share conceptual contexts with one another would therefore come to have partially shared frontal representations. This shared representation, representing only shared conceptual context, is a purely abstract representation. As a simple example, the abstract concept 'better' might be represented by a positive affective value as well as a predictive representation of an attentional shift to a different representation associated with a negative affective value. Once abstract representations have been extracted in this way, the frontal lobe's reinforcement driven mechanisms for gating and sustaining activation could be employed to combine multiple representations into a common schema, run perceptual simulations (cf. Barsalou et al, 2003) to project the ramifications of these new combined representations and expand their repertoires of contextual relationships, and manipulate them in other ways.

Clearly this is a speculative framework. It has yet to be modeled or empirically tested. But it is an example of one way that a cortical model integrating CLA-like functionality could offer new approaches to a longstanding problem. It may be that the same mechanism that in posterior cortex produces representations of object identity that are invariant to visual transformation, is employed by frontal cortex to produce representations of contextual relationships that are invariant to object identity.

## Conclusions

While there is an intriguing body of evidence pointing to the possibility of its existence, any attempt to identify a common algorithm that characterizes the processing carried out by all areas of cortex must confront the great variety of functions in which the cortex participates. In this review, I have singled out the HTM CLA as a model of cortical function that is compatible with biological data and that encompasses many properties associated with cortical function, including the use of sparse distributed representations, hierarchical Bayesian inference, and predictive coding. Perhaps most distinctively, the CLA offers an explanation for how representations, at higher levels, become increasingly invariant to rapidly changing stimulus attributes. By exploiting the principle of temporal slowness to solve this problem, the CLA provides an alternative to the hard-coded connections or supervised error-driven learning employed by other systems to generate visual representations invariant to transformations of position, scale, rotation, pose, and so on, and also suggests a process by which abstract representations of contextual relationship roles may be generated in frontal cortex.

If the HTM CLA is to be considered a candidate for a universal cortical algorithm, it must be well suited not only to the requirements of the posterior cortical functions to which the HTM models have most often been applied, but also to the requirements of frontal cortical function. By examining the nature of the close interactions between frontal cortex and subcortical structures, it is possible to separate out the set of functions that the cortex itself may be responsible for, from those functions that emerge from subcortical interactions. A relatively simple model of the interactions between frontal cortex and the cortico-thalamic relay loops controlled by the basal ganglia reinforcement learning system (Frank & Badre, 2012) demonstrates that interactions between frontal areas and subcortical structures are capable of explaining a wide range of control functions attributed to frontal cortex, including the hierarchical selection of both strategy and response, top-down attentional effects, and the updating of actively maintained working memory representations. These results suggest that the only function intrinsic to cortex that is necessary to accommodate frontal functions, in addition to those also required to accommodate posterior functions, is the ability to support the active maintenance of representations.

Typically connectionist models of frontal cortical function have been weakest in terms of the richness and versatility of their representations. Besides providing for competitive Hebbian learning which mediates the representation of groups of commonly co-occurring inputs, the CLA offers two additional forms of representation that both take advantage of its ability to predictively activate representations of expected future inputs. First, it learns repeated sequences of inputs and re-represents them as static patterns at higher levels, which allows for the sort of hierarchical compression and expansion of time that is well suited to both perceptual and executive processing. Second, CLA representations that, as a set, have a high probability of alternating among one another in being active, come to incorporate predictive sub-representations of one another, and so come to be partially overlapping. This effect provides for perceptual invariance, may serve to unite related perceptual and conceptual representations into common schemas, and may facilitate the extraction of abstract representations. In these ways, the principles underlying the CLA may offer a path toward providing the representational strengths that have so far eluded connectionist models of frontal cortical function.

The HTM CLA is early in development and has yet to be widely researched, applied, or empirically tested. I do not propose that the CLA in its current form is a complete or accurate realization of a universal cortical algorithm. I do suggest, however, that the principles embodied by the CLA offer valuable insights toward the solutions of a number of longstanding problems regarding cortical function. There are many clear opportunities for extending the power, biological grounding and practical applicability of the CLA, such as in the integration of important insights from complementary biologically-based cortical models and the coupling of the CLA with models of subcortical functions. For these reasons I believe that further research using the CLA stands to yield valuable results.

# References

Alexander, G.E., DeLong, M.R., & Strick, P.L. (1986). Parallel organization of functionally segregated circuits linking basal ganglia and cortex. *Annual Review of Neuroscience*, 9, 357–381.

Alink, A., Schwiedrzik, C.M., Kohler, A., Singer, W., & Muckli, L. (2010). Stimulus predictability reduces responses in primary visual cortex. *Journal of Neuroscience*, 30, 2960–2966.

Amodio, D.M., Frith, C.D. (2006). Meeting of minds: the medial frontal cortex and social cognition. *Nat Rev Neurosci*, 7, 268 –277.

Anderson, J. R. (1983). The architecture of cognition. Cambridge, MA: Harvard University Press.

Asaad, W.F., Rainer, G., Miller, E.K. (1998). Neural activity in the primate prefrontal cortex during associative learning. *Neuron*, 21, 1399–407.

Asaad, W.F., Rainer, G., & Miller, E.K. (2000). Task specific neural activity in the primate prefrontal cortex. *J. Neurophysiol.*, 84, 451–59.

Atkinson, R.C., Shiffrin, R.M. (1968). Human memory: A proposed system and its control processes. In Spence, K.W., Spence, J.T. eds. *The psychology of learning and motivation (Volume 2)*. New York: Academic Press. pp. 89–195.

Baddeley, A.D., & Hitch, G.J. (1974). Working memory. In G.A. Bower (Ed.), *Recent advances in learning and motivation* (Vol. 8, pp. 47–90). New York: Academic Press.

Baddeley, A.D. (1986). *Working Memory*. Oxford: Oxford University Press.

Baddeley, A.D. (2000). The episodic buffer: A new component of working memory? *Trends in Cognitive Sciences*, 4(11), 417-423.

Badre, D., Kayser, A., D'Esposito, M. (2010). Frontal cortex and the discovery of abstract action rules. *Neuron*, 66, 315–326.

Bair, W., Cavanaugh, J.R., & Movshon, J.A. (2003). Time course and time-distance relationships for surround suppression in macaque V1 neurons. *Journal of Neuroscience*, 23, 7690–7701.

Bar, M. (2009). The proactive brain: memory for predictions. *Phil. Trans. R. Soc. B*, 364(1521), 1235-1243.

Barbas, H., & Pandya, D.N. (1989). Architecture and intrinsic connections of the prefrontal cortex in the rhesus monkey. *Journal of Comparative Neurology*, 286(3), 353–375.

Barsalou, L.W., Kyle Simmons, W., Barbey, A.K., Wilson, C.D. (2003). Grounding conceptual knowledge in modality-specific systems. *Trends Cogn. Sci.* 7, 84–91.

Banich, M.T., Milham, M.P., Atchley, R., Cohen, N.J., Webb A, et al. (2000). Prefrontal regions play a predominant role in imposing an attentional "set": evidence from fMRI. *Cogn. Brain Res.*, 10, 1–9.

Baker, S.C, Rogers, R.D., Owen, A.M., Frith, C.D., Dolan, R.J., Frackowiak, R.S.J., & Robbins, T.W. (1996). Neural systems engaged by planning: A Pet study of the Tower of London task. *Neuropsychologia*, 34, 515.

Barbas, H. (2006). Organization of the principal pathways of prefrontal lateral, medial, and orbitofrontal cortices in primates and implications for their collaborative interaction in executive functions. *The frontal lobes. Development, function and pathology*, 21-68. Cambridge University Press.

Bastos, A.M., Usrey, W.M., Adams, R.A., Mangun, G.R., Fries, P., Friston, K.J. (2012). Canonical microcircuits for predictive coding. *Neuron*, 76, 695–711.

Bedny, M., Dodell-Feder, D., Pascual-Leone, A., Fedorenko, E., & Saxe, R. (2011). Language processing in the occipital cortex of congenitally blind adults. *Proceedings of the National Academy of Sciences of the United States of America*, 108(11), 4429–4434.

Bengio, Y. (2009). Learning deep architectures for AI. *Foundations and Trends in Machine Learning*, 2(1).

Berkes, P. (2005). Temporal slowness as an unsupervised learning principle: self-organization of complex-cell receptive fields and application to pattern recognition. Ph.D. dissertation, Humboldt University, Berlin.

Bienenstock, E.L., Cooper, L.N., & Munro, P.W. (1982). Theory for the development of neuron selectivity: orientation specificity and binocular interaction in visual cortex. *Journal of Neuroscience*, 2(2), 32–48.

Botvinick, M.M., Braver, T.S., Barch, D.M., Carter, C.S. & Cohen, J.D. (2001). Conflict monitoring and cognitive control. *Psychol Rev*, 108, 624–652.

Braver, T.S., Cohen, J.D., eds. (2000). *On the Control of Control: The Role of Dopamine in Regulating Prefrontal Function and Working Memory*. Cambridge, MA: MIT Press.

Brown, J.W., Bullock, D., & Grossberg, S. (2004). How laminar frontal cortex and basal ganglia circuits interact to control planned and reactive saccades. *Neural Networks*, 17(4), 471-510.

Bunge, S. A. (2004). How we use rules to select actions: a review of evidence from cognitive neuroscience. *Cognitive, Affective, & Behavioral Neuroscience*, 4(4), 564-579.

Buonomano, D.V., Karmarkar, U.R. (2002). How do we tell time? *Neuroscientist* 8, 42-51.

Bussey, T. J., Wise, S. P., & Murray, E. A. (2002). Interaction of ventral and orbital prefrontal cortex with inferotemporal cortex in conditional visuomotor learning. *Behavioral Neuroscience*, 116, 703-715.

Buzo, A., Gray Jr, A., Gray, R., & Markel, J. (1980). Speech coding based upon vector quantization. Acoustics, Speech and Signal Processing, *IEEE Transactions on*, 28(5), 562-574.

Callaway, E.M. (1998). Local circuits in primary visual cortex of the macaque monkey. *Annual Review of Neuroscience*. 21, 47–74.

Carpenter, G. & Grossberg, S. A massively parallel architecture for a self-organizing neural pattern recognition. *Machine. Comp. Vision, Graphics and Image Proc*. 37, 54-115.

Chao, L.L., Knight, R.T. (1997). Prefrontal deficits in attention and inhibitory control with aging. *Cereb. Cortex*, 7, 63–9.

Chellapilla, K., Puri, S. & Simard, P. (2006). High performance convolutional neural networks for document processing. In *International Workshop on Frontiers in Handwriting Recognition.*

Chevalier, G., & Deniau, J.M. (1990). Disinhibition as a basic process in the expression of striatal functions. *Trends in neurosciences*, 13(7), 277-280.

Christoff, K., Keramatian, K., Gordon, A.M., Smith, R. & Madler, B. (2009). Prefrontal organization of cognitive control according to levels of abstraction. *Brain Res.*, 1286, 94–105.

Churchland, P.S., Sejnowski, T.J. (1992). *The Computational Brain*. MIT Press, Cambridge, MA.

Ciresan, D., Meier, U., Masci, J., Gambardella, L. & Schmidhuber, J. (2011). Flexible, high performance convolutional neural networks for image classification. In *International Joint Conference on Artificial Intelligence*, pages 1237-1242.

Ciresan, D., Meier, U., Masci, J., & Schmidhuber, J. (2012). Multi-column deep neural network for traffic sign classification. *Neural Networks*, 32, 333-338.

Ciresan, D., Giusti, A., Gambardella, L., & Schmidhuber, J. (2012). Deep neural networks segment neuronal membranes in electron microscopy images. *Advances in Neural Information Processing Systems*, Lake Tahoe.

Ciresan, D., Meier, U., Schmidhuber, J. (2012). Multi-column deep neural networks for image classification. *IEEE Conf. on Computer Vision and Pattern Recognition (CVPR 2012)*.

Cohen, J.D., Braver, T.S., O'Reilly, R.C. (1996). A computational approach to prefrontal cortex, cognitive control, and schizophrenia: Recent developments and current challenges. *Philos. Trans. Roy. Soc. London B*. 351, 1515–1527.

Cohen, J.D., Barch, D.M., Carter, C.S., Servan-Schreiber, D. (1999). Schizophrenic deficits in the processing of context: converging evidence from three theoretically motivated cognitive tasks. *Journal of Abnormal Psychology.* 108, 120–33.

Colby, C.L., Duhamel, J.R., & Goldberg, M.E. (1996). Visual, presaccadic, and cognitive activation of single neurons in monkey lateral intraparietal area. *Journal of neurophysiology*, 76(5), 2841-2852.

Constantinidis, C., Franowicz, M. N., & Goldman-Rakic, P. S. (2001). The sensory nature of mnemonic representation in the primate prefrontal cortex. *Nature Neuroscience*, 4, 311-316.

Corbetta, M., Akbudak, E., Conturo, T.E., Snyder, A.Z., Ollinger, J.M., et al. (1998). A common network of functional areas for attention and eye movements. *Neuron*, 21, 761–73.

Corbetta, M., Miezin, F.M., Shulman, G.L., Petersen, S.E. (1993). A PET study of visuospatial attention. *J. Neurosci.*, 13, 1202–26.

Cox, D.D., Meier, P., Oertelt, N., & DiCarlo, J. (2005). 'Breaking' position-invariant object recognition. *Nature Neuroscience*, 8, 1145–1147.

Cromwell, H.C., & Berridge, K.C. (1996). Implementation of action sequences by a neostriatal site: a lesion mapping study of grooming syntax. *The Journal of neuroscience*, 16(10), 3444-3458.

DeFelipe, J., Alonso-Nanclares, L., Arellano, J.I. (2002). Microstructure of the neocortex: Comparative aspects. *J Neurocytol*, 31, 299–316.

Desimone, R. & Duncan, J. (1995). Neural mechanisms of selective attention. *Annual Review of Neuroscience*, 18, 193-222.

Desmurget, M., & Turner, R.S. (2010). Motor sequences and the basal ganglia: kinematics, not habits. *The Journal of Neuroscience*, 30(22), 7685-7690.

D'Esposito, M., Ballard, D., Zarahn, E., & Aguirre, G. K. (2000). The role of prefrontal cortex in sensory memory and motor preparation: An event-related fMRI study. *NeuroImage*, 11, 400-408.

de-Wit, L., Machilsen, B., & Putzeys, T. (2010). Predictive coding and the neural response to predictable stimuli. *Journal of Neuroscience*, 30(26), 8702-8703.

Diamond, A., & Goldman-Rakic, P. S. (1989). Comparison of human infants and rhesus monkeys on Piaget's AB task: Evidence for dependence on dorsolateral prefrontal cortex. *Experimental Brain Research*, 74, 24-40.

Dias, R., Robbins, T.W., Roberts, A.C. (1996). Primate analogue of the Wisconsin Card Sorting Test: effects of excitotoxic lesions of the prefrontal cortex in the marmoset. *Behav. Neurosci.*, 110, 872–86.

DiCarlo, J. & Cox, D. (2007). Untangling invariant object recognition. *Trends in Cognitive Sciences*, 11(8), 333-341.

Dunbar K, Sussman D. (1995). Toward a cognitive account of frontal lobe function: simulating frontal lobe deficits in normal subjects. *Annals of the New York Academy of Sciences*, 769, 289–304.

Duncan, J. (1986). Disorganization of behaviour after frontal lobe damage. *Cognitive Neuropsychology*, 3, 271–90.

Etkin, A., Egner, T., Peraza, D.M., Kandel, E.R. & Hirsch, J. (2006). Resolving emotional conflict: a role for the rostral anterior cingulate cortex in modulating activity in the amygdala. *Neuron*, 51, 871– 882.

Feldman, D.E. (2009). Synaptic mechanisms for plasticity in neocortex. *Annual Review of Neuroscience*. 32, 33–55

Felleman, D.J. & Van Essen, D.C. (1991). Distributed hierarchical processing in the primate cerebral cortex. *Cerebral Cortex*, 1, 1-47.

Ferrier, D. (1874). Experiments on the brain of monkeys - No. 1. *Proc. R. Soc. Lond*, 23 (156–163), 409–430.

Ferrier, M. (2006). *From Vision to Language: A Domain General Approach to Statistical Learning.* Unpublished manuscript.

Field, D.J. (1987). Relations between the statistics of natural images and the response properties of cortical cells. *Journal of the Optical Society of America A*, 4, 2379-2394.

Fink, G.R., Dolan, R.J., Halligan, P.W., Marshall, J.C., Frith, C.D. (1997). Space-based and object based visual attention: shared and specific neural domains. *Brain*, 120, 2013–28.

Fiser, J., & Aslin, R. N. (2002). Statistical learning of new visual feature combinations by infants. *Proceedings of the National Academy of Science*, 99, 15822-15826.

Földiák, P. (1991). Learning invariance from transformation sequences. *Neural Computation*, 3, 194–200.

Földiák, P. (2002). Sparse coding in the primate cortex. In M.A. Arbib (Ed.), *The Handbook of Brain Theory and Neural Networks (2nd ed.)* (pp. 1064-1068). Cambridge, MA: MIT Press.

Forbes, C.E., & Grafman, J. (2010). The Role of the Human Prefrontal Cortex in Social Cognition and Moral Judgment. *Annual review of neuroscience*, 33, 299-324.

Frank, M.J., & Badre, D. (2012). Mechanisms of hierarchical reinforcement learning in corticostriatal circuits 1: computational analysis. *Cerebral cortex*, 22(3), 509-526.

Frank, M.J., Loughry, B., & O'Reilly, R.C. (2001). Interactions between frontal cortex and basal ganglia in working memory: a computational model. *Cognitive, Affective, & Behavioral Neuroscience*, 1(2), 137-160.

Frank, M.J. & O'Reilly, R.C. (2006). A mechanistic account of striatal dopamine function in human cognition: Psychopharmacological studies with cabergoline and haloperidol. *Behavioral Neuroscience*, 120, 497-517.

Frank, M.J., Santamaria, A., O'Reilly, R. & Willcutt, E. (2007). Testing computational models of dopamine and noradrenaline dysfunction in Attention Deficit/Hyperactivity Disorder. *Neuropsychopharmacology*, 32, 1583-99.

Frank, M.J., Seeberger, L. & O'Reilly, R.C. (2004). By carrot or by stick: Cognitive reinforcement learning in Parkinsonism. *Science*, 306, 1940-1943.

Friston, K. (2005). A theory of cortical responses. *Philos. Trans. R. Soc. Lond. B: Biol. Sci.*, 360, 815–836

Fukushima, K. (1988). Neocognitron: A Hierarchical Neural Network Capable of Visual Pattern Recognition. *Neural Networks*, 1(2), 119-130.

Fuster, J.M. (1973). Unit activity in prefrontal cortex during delayed-response performance: neuronal correlates of transient memory. *J. Neurophysiol*, 36, 61–78.

Fuster, J.M. (1980). *The prefrontal cortex*. Ney York: Raven Press.

Fuster, J.M. (1997). *The prefrontal cortex*, 3rd edn. New York: Lippincott-Raven.

Fuster, J.M. (1999). Cognitive functions of the frontal lobes. In Miller, B.L., Cummings, J.L., eds. *The human frontal lobes: Functions and disorders.* New York, NY, US: Guilford Press. pp. 187-195.

Fuster, J.M., Bauer, R.H., Jervey, J.P. (1982). Cellular discharge in the dorsolateral prefrontal cortex of the monkey in cognitive tasks. *Exp. Neurol.,* 77, 679–94.

Garrido, M.I., Kilner, J.M., Kiebel, S.J., & Friston, K.J. (2007). Evoked brain responses are generated by feedback loops. *Proc. Natl. Acad. Sci. USA* 104, 20961–20966.

Garrido, M.I., Kilner, J.M., Stephan, K.E., & Friston, K.J. (2009). The mismatch negativity: a review of underlying mechanisms. *Clin. Neurophysiol.* 120, 453–463.

Gehring, W.J. & Willoughby, A.R. (2002). The medial frontal cortex and the rapid processing of monetary gains and losses. *Science*, 295, 2279 –2282.

George, D. (2008). How the brain might work: A hierarchical and temporal model for learning and recognition. Ph.D. Thesis, Stanford University.

George, D., Hawkins, J. (2005). A hierarchical Bayesian model of invariant pattern recognition in the visual cortex. *Proceedings of the International Joint Conference on Neural Networks*, 3, 1812–1817.

George, D., Hawkins, J. (2009). Towards a mathematical theory of cortical micro-circuits. *PLoS Comput. Biol.*, 5(10).

Gershberg, F.B., Shimamura, A.P. (1995). Impaired use of organizational strategies in free recall following frontal lobe damage. *Neuropsychologia*, 13, 1305–33.

Goldman-Rakic, P.S. (1987). Circuitry of primate prefrontal cortex and regulation of behavior by representational memory. In *Handbook of physiology, the nervous system, higher functions of the brain* (ed. F. Plum), sect. I, vol. V, pp. 373-417. Bethesda, MD: American Physiological Society.

Goldman-Rakic, P.S., Cools, A.R., & Srivastava, K. (1996). The prefrontal landscape: implications of functional architecture for understanding human mentation and the central executive [and Discussion]. *Philosophical Transactions of the Royal Society of London. Series B: Biological Sciences*, 351(1346), 1445-1453.

Graf Estes, K., Evans, J. L., Alibali, M. W., Saffran, J. R. (2007). Can infants map meaning to newly segmented words? *Psychological Science*, 18, 254-259.

Grafman, J. (2002). The structured event complex and the human prefrontal cortex. In *Principles of Frontal Lobe Function*, ed. D.T.H. Stuss, R.T. Knight, p. 616. Oxford/New York: Oxford Univ. Press.

Granger, R. (2006). Engines of the brain: The computational instruction set of human cognition. *AI Magazine*, 27(2), 15.

Graybiel, A.M. (1995). Building action repertoires: memory and learning functions of the basal ganglia. *Current opinion in neurobiology*, 5(6), 733-741.

Grenander, U. (1993). *General Pattern Theory*. Oxford University Press, Oxford.

Gross, C.G., Rocha-Miranda, C., and Bender, D. (1972). Visual properties of neurons in the inferotemporal cortex of the macaque. *Journal of Neurophysiology*, 35, 96-111.

Grossberg, S. (2007). Towards a unified theory of neocortex: laminar cortical circuits for vision and cognition. *Progress in Brain Research*. 165, 79–104.

Guitton, D., Buchtel, H.A., & Douglas, R.M. (1985). Frontal lobe lesions in man cause difficulties in suppressing reflexive glances and in generating goal-directed saccades. *Experimental Brain Research*, 58, 455-472.

Halsband, U., Passingham, R.E. (1985). Premotor cortex and the conditions for movement in monkeys. *Behav. Brain Res.*, 18, 269–76.

Hawkins, J., Ahmad, S., & Dubinsky, D. (2010). Hierarchical temporal memory including HTM cortical learning algorithms. Technical report, Numenta, CA.

Hawkins, J. (2004). *On Intelligence*. New York: Henry Holt.

Hazy, T.E., Pauli, W., Herd, S., others, & O'Reilly, R.C. (in preparation). Neural mechanisms of executive function: Biological substrates of active maintenance and adaptive updating.

Hecht-Nielsen, R. (2007). *Confabulation Theory*. Heidelberg: Springer-Verlag.

Herculano-Housel, S., Collins, C.E., Wang, P., Kaas, J. (2008). The basic nonuniformity of the cerebral cortex. *Proc Natl Acad Sci USA*, 105, 12593–12598.

Hinton, G. E., McClelland, J. L., & Rumelhart, D. E. (1986). Distributed representations. In D. E. Rumelhart, J. L. McClelland, & the PDP Research Group (Eds.), *Parallel distributed processing: explorations in the microstructure of cognition: Vol. 1. Foundations* (pp. 77-109). Cambridge, MA: MIT Press.

Hinton, G. E., Osindero, S., and Teh, Y. W. (2006). A fast learning algorithm for deep belief nets. *Neural Computation*, 18, 1527-1554.

Hirsch, J.A. & Martinez, L. M. (2006). Laminar processing in the visual cortical column. *Current Opinion in Neurobiology*, 16, 377–384.

Hubel, D.H. and Wiesel, T.N. (1962). Receptive fields, binocular interaction and functional architecture in the cat's visual cortex. *Journal of Physiology*, 160, 106–154.

Hubel, D.H., & Wiesel, T.N. (1968). Receptive fields and functional architecture of monkey striate cortex. *The Journal of Physiology*, 195, 215-243.

Hupe´, J.M., James, A.C., Payne, B.R., Lomber, S.G., Girard, P., & Bullier, J. (1998). Cortical feedback improves discrimination between figure and background by V1, V2 and V3 neurons. *Nature* 394, 784–787.

Jackendoff, R. (2002). *Foundations of Language: Brain, Meaning, Grammar, Evolution.* Oxford: Oxford University Press.

Janowsky, J.S., Shimamura, A.P., Kritchevsky, M., Squire, L.R. (1989). Cognitive impairment following frontal lobe damage and its relevance to human amnesia. *Behavioral Neuroscience*, 103, 548–60.

Jilk, D.J., Lebiere, C., O'Reilly, R.C., & Anderson, J.R. (2008). SAL: An explicitly pluralistic cognitive architecture. *Journal of Experimental and Theoretical Artificial Intelligence*, 20(3), 197-218.

Jin, D.Z., Fujii, N., Graybiel, A.M. (2009) Neural representation of time in cortico-basal ganglia circuits. *Proceedings of the National Academy of Sciences*, 106, 19156–19161.

Kastner, S., De Weerd, P., Elizondo, I., Desimone, R., Ungerleider, L.G. (1998). Mechanisms of spatial attention in human extrastriate cortex as revealed by functional MRI. *Soc. Neurosci. Abstr.*, 24, 1249.

Kastner, S., Ungerleider, L.G. (2000). Mechanisms of visual attention in the human cortex. *Annu. Rev. Neurosci.*, 23, 315–41.

Kaufer, D.I., & Lewis, D.A. (1999). Frontal lobe anatomy and cortical connectivity. *The human frontal lobes*, 27-44. New York: Guilford Press.

Kirkham, N. Z., Slemmer, J., A., & Johnson, S. P. (2002). Visual statistical learning in infancy: evidence for a domain general learning mechanism. *Cognition*, 83, B35-B42.

Kodaka, Y., Mikami, A., Kubota, K. (1997). Neuronal activity in the frontal eye field of the monkey is modulated while attention is focused onto a stimulus in the peripheral visual field, irrespective of eye movement. *Neurosci. Res.*, 28, 291–98.

Koechlin, E., Corrado, G., Pietrini, P., Grafman, J. (2000). Dissociating the role of the medial and lateral anterior prefrontal cortex in human planning. *Proc Natl Acad Sci U S A*. 97,7651–7656.

Koechlin, E., Ody, C., Kouneiher, F. (2003). The architecture of cognitive control in the human prefrontal cortex. *Science*. 302, 1181–1184.

Kohonen, T. (1982). Self-organized formation of topologically correct feature maps. *Biological Cybernetics*, 43, 59-69.

Kok, P., Jehee, J., & de Lange, P. (2012). Less is more: expectation sharpens representations in the primary visual cortex.

Kritzer, M.F., Goldman-Rakic, P.S. (1995). Intrinsic circuit organization of the major layers and sublayers of the dorsolateral prefrontal cortex in the rhesus monkey. *J Comp Neurol*, 359, 131–143.

Leichnetz, G.R., & Astruc, J. (1975). Preliminary evidence for a direct projection of the prefrontal cortex to the hippocampus in the squirrel monkey. *Brain, Behavior and Evolution*, 11(5-6), 355-364.

Lennie, P. (2003). The cost of cortical computation. *Current Biology*, 13:493-497.

Leon, M.I., Shadlen, M.N. (1999). Effect of expected reward magnitude on the response of neurons in the dorsolateral prefrontal cortex of the macaque. *Neuron*, 24, 415–25.

Levitt, J.B., Lewis, D.A., Yoshioka, T., & Lund, J.S. (1993). Topography of pyramidal neuron intrinsic connections in macaque monkey prefrontal cortex (areas 9 and 46). *Journal of Comparative Neurology*, 338(3), 360-376.

Lhermitte, F. (1983). "Utilization behaviour" and its relation to lesions of the frontal lobes. *Brain*, 106, 237–55.

Li, N., & DiCarlo, J. J. (2008). Unsupervised natural experience rapidly alters invariant object representation in visual cortex. *Science*, 321(5895), 1502-1507.

Logothetis, N.K. (1998). Object vision and visual awareness. *Current Opinions in Neurobiology*, 8(4), 536-44.

Lucke, J. and Bouecke, J.D. (2005). Dynamics of cortical columns – self-organization of receptive fields. *Proceedings of the International Conference on Artificial Neural Networks*. LNCS 3696:31–37.

Luria, A.R., Karpv, B.A., & Yarbuss, A.L. (1966). Disturbances of active visual perception with lesions of the frontal lobes. *Cortex*, 2, 202-212.

Luria, A.R. (1969). Frontal lobe syndromes. In *Handbook of Clinical Neurology*, ed. P.J. Vinken, G.W. Bruyn, pp. 725–57. New York: Elsevier.

Maltoni, D. (2011). Pattern recognition by hierarchical temporal memory. Technical report, DEIS University of Bologna.

Markram H., Perin R. (2011). Innate neural assemblies for lego memory. *Front. Neural Circuits* 5:6. doi: 10.3389/fncir.2011.00006.

Marr, D. (1982). *Vision. A Computational Investigation into the Human Representation and Processing of Visual Information*. W.H. Freeman, San Francisco.

Martinez, L.M., Wang, Q., Reid, R.C., Pillai, C., Alonso, J.M., Sommer, F.T., Hirsch, J.A. (2005). Receptive field structure varies with layer in the primary visual cortex. *Nature Neuroscience*, 8, 372-379.

Masquelier, T., Serre, T., Thorpe, S., & Poggio, T. (2007). Learning Complex Cell Invariance from Natural Videos: A Plausibility Proof. *MIT Center for Biological & Computational Learning Paper #269*/Mit-csaiLtr #2007-060, Cambridge, MA.

McClelland, J.L. (1994). The interaction of nature and nurture in development: A parallel distributed processing perspective. *International perspectives on psychological science*, 1, 57-88.

McClelland, J.L & Rumelhart, D.E. (1981). An interactive activation model of context effects in letter perception. Part I: an account of basic findings. *Psychol. Review*, 88, 375-407.

McClelland, J.L., Rumelhart, D.E., & the PDP Research Group (Eds.). (1986). *Parallel Distributed Processing: Explorations in the Microstructure of Cognition , Vol. 2: Psychological and Biological Models*. MIT Press.

Mel, B. (1999). Why have dendrites? A computational perspective. *Dendrites*. Oxford University Press, Oxford.

Mesulam, M.M. (1997). Anatomic principles in behavioral neurology and neuropsychology. Feinberg, T.E. & Farah, M.J. (Eds.), *Behavioral neurology and neuropsychology*, 55-68. New York: McGraw-Hill.

Miller, E.K., Erickson, C.A., Desimone, R. (1996). Neural mechanisms of visual working memory in prefrontal cortex of the macaque. *J. Neurosci.*, 16, 5154–67.

Miller, E.K., & Cohen, J.D. (2001). An integrative theory of prefrontal cortex function. *Annual review of neuroscience*, 24(1), 167-202.

Milner B. (1963). Effects of different brain lesions on card sorting. *Arch. Neurol*, 9:90.

Mitchison, G. (1991). Removing time variation with the anti-Hebbian differential synapse. *Neural Computation*, 3, 312–320.

Miyashita, Y. (1988). Neuronal correlate of visual associate long-term memory in the primate temporal cortex. *Nature*, 335, 817–820.

Mountcastle, V.B. (1978). An organizing principle for cerebral function: the unit model and the distributed system. In G.M. Edelman and V.B. Mountcastle (Eds.), *The Mindful Brain*. Cambridge, MA: MIT Press

Mountcastle, V.B. (1997). The columnar organization of the neocortex. *Brain*, 120, 701-722.

Moustafa, A.A., Sherman, S.J. & Frank, M.J. (2008). A dopaminergic basis for working memory, learning and attentional shifting in Parkinsonism. *Neuropsychologia*, 46, 3144-3156.

Mumford, D. (1992). On the computational architecture of the neocortex II. *Biological Cybernetics*, 66, 241-251.

Nagel, I.E., Schumacher, E.H., Goebel, R., & D'Esposito, M. (2008). Functional MRI investigation of verbal selection mechanisms in lateral prefrontal cortex. *Neuroimage*, 43(4), 801-807.

Nauta, W.J.H. (1964). Some efferent connections of the prefrontal cortex in the monkey. *The frontal granular cortex and behavior*. New York: McGraw-Hill, pp. 397-409.

Newell, A. (1973). 'Production systems: Models of control structures', in *Visual Information Processing*, ed. W.G. Chase, New York, NY: Academic Press, pp. 463–525.

Norman, D.A., & Shallice, T. (1980). Attention to action: Willed and automatic control of behavior. California Univ. San Diego La Jolla Center for Human Information Processing.

Norman, D.A., & Shallice, T. (1986). Attention to action: Willed and automatic control of behavior. In *Consciousness and self-regulation* (ed. G.E. Schwartz & D. Shapiro), vol 4. Plenum Press: New York.

O'Doherty, J., Rolls, E.T., Francis, S., Bowtell, R., McGlone, F. et al. (2000). Sensory-specific satiety-related olfactory activation of the human orbitofrontal cortex. *NeuroReport*, 11, 893–97.

Olshausen, B.A., Field, D.J. (1996). Emergence of simple cell receptive field properties by learning a sparse code for natural images. *Nature*, 381, 607-609.

Olshausen, B. & Field, D. (2004). Sparse coding of sensory inputs. *Current Opinion in Neurobiology*, 14, 481-487.

O'Reilly, R.C. (1996). Biologically plausible error-driven learning using local activation differences: The generalized recirculation algorithm. *Neural computation*, 8(5), 895-938.

O'Reilly, R.C. (2010). The What and How of prefrontal cortical organization. *Trends in neurosciences*, 33(8), 355-361.

O'Reilly, R.C., Frank, M.J., Hazy, T.E., & Watz, B. (2007). PVLV: the primary value and learned value Pavlovian learning algorithm. *Behavioral neuroscience*, 121(1), 31.

O'Reilly, R. & Munakata, Y. (2000). Computational explorations in cognitive neuroscience. MIT Press, Cambridge Massachusetts.

O'Reilly, R.C., Munakata, Y., Frank, M.J., Hazy, T.E., & Contributors (2012). *Computational Cognitive Neuroscience*. Wiki Book, 1st Edition, URL: http://ccnbook.colorado.edu.

O'Reilly, R. C., Hazy, T. E., & Herd, S. A. (2012). The Leabra Cognitive Architecture: How to Play 20 Principles with Nature and Win! In S. Chipman (Ed) *Oxford Handbook of Cognitive Science*, Oxford: Oxford University Press.

Pearl, J. (1988). *Probabilistic reasoning in intelligent systems: networks of plausible inference*. Morgan Kaufmann, San Francisco.

Perret, E. (1974). The left frontal lobe of man and the suppression of habitual responses in verbal categorical behaviour. *Neuropsychologia*, 12, 323–30.

Perrett, D.I., Rolls, E.T., Caan, W. (1982) Visual neurons responsive to faces in the monkey temporal cortex. *Experimental Brain Research*, 47(3), 329-342.

Petrides, M. (1985). Deficits in non-spatial conditional associative learning after periarcuate lesions in the monkey. *Behav. Brain Res.*, 16, 95–101.

Petrides, M. (1990). Nonspatial conditional learning impaired in patients with unilateral frontal but not unilateral temporal lobe excisions. *Neuropsychologia*, 28, 137–49.

Pinker, S. (1994). *The language instinct: How the mind creates language*, pp. 37–43. W. Morrow, New York.

Pribram, K.H. (1987). The subdivisions of the frontal cortex revisited. In *The frontal lobes revisited* (ed. E. Perecman), pp. 11-39. New York: The IRBN Press.

Price, J.L. (1999). Prefrontal cortical networks related to visceral function and mood. *Ann. NY Acad. Sci.*, 877, 383–96.

Price, R. W. (2011). Hierarchical temporal memory cortical learning algorithm for pattern recognition on multi-core architectures. M.Sc. thesis, Portland State University.

Ranzato, M. Huang, F., Boureau, Y., LeCun, Y. (2007). Unsupervised Learning of Invariant Feature Hierarchies with Applications to Object Recognition. *Proc. Computer Vision and Pattern Recognition, 2007*.

Rao, R. & Ballard, D. (1997). Dynamic model of visual recognition predicts neural response properties in the visual cortex. *Neural Computation*, 9, 721-763.

Rao, R.P., and Ballard, D.H. (1999). Predictive coding in the visual cortex: a functional interpretation of some extra-classical receptive-field effects. *Nat. Neurosci.*, 2, 79–87.

Rao, S.G., Williams, G.V., & Goldman-Rakic, P.S. (1999). Isodirectional tuning of adjacent interneurons and pyramidal cells during working memory: evidence for microcolumnar organization in PFC. *Journal of Neurophysiology*, 81(4), 1903-1916.

Redgrave, P., Prescott, T.J., & Gurney, K. (1999). The basal ganglia: a vertebrate solution to the selection problem?. *Neuroscience*, 89(4), 1009-1023.

Rescorla, R.A., & Wagner, A.R. (1972). A theory of Pavlovian conditioning: Variation in the effectiveness of reinforcement and nonreinforcement. In A. H. Black & W. F. Prokasy (Eds.), *Classical conditioning II: Theory and research* (pp. 64–99). New York: Appleton-Century-Crofts.

Riesenhuber, M. & Poggio, T. (1999). Hierarchical models of object recognition in cortex. *Nat Neurosci*, 2, 1019–1025.

Rockel, A.J., Hiorns, R.W., Powell, T.P.S. (1980). The basic uniformity in structure of the neocortex. *Brain* 103, 221–244.

Rodriguez, A., Whitson, J., Granger, R. (2005). Derivation and analysis of basic computational operations of thalamocortical circuits. *Journal of Cognitive Neuroscience*, 16, 856-877.

Rosene, D.L., & Van Hoesen, G.W. (1977). Hippocampal efferents reach widespread areas of cerebral cortex and amygdala in the rhesus monkey. *Science*, 198(4314), 315-317.

Rossi, A.F., Rotter, P.S., Desimone, R., & Ungerleider, L.G. (1999). Prefrontal lesions produce impairments in feature-cued attention. *Soc. Neurosci. Abst.,* 25, 3.

Rumelhart, D.E., Hinton, G.E., & Williams, R.J. (1986). Learning representations by back-propagating errors. *Nature*, 323(6088), 533–536.

Rumelhart, D.E., McClelland, J.L., & the PDP Research Group (Eds.). (1986b). *Parallel Distributed Processing: Explorations in the Microstructure of Cognition, Vol.1: Foundations , Vol. 1: Foundations*. Cambridge, MA: MIT Press.

Rumelhart, D.E., & Zipser, D. (1985). Feature discovery by competitive learning. *Cognitive science*, 9(1), 75-112.

Rushworth, M.F., Behrens, T.E. (2008). Choice, uncertainty and value in prefrontal and cingulate cortex. *Nat Neurosci*, 11, 389 –397.

Saffran, J. R., Newport, E. L., & Aslin, R. N. (1996a). Word segmentation: the role of distributional cues. *Journal of Memory and Language*, 35, 606-621.

Saffran, J. R., Aslin, R. N., & Newport, E. L. (1996b). Statistical learning by 8-month-old infants. *Science*, 274, 1926-1928.

Saffran, J. R., Johnson, E. K., Aslin, R. N. & Newport, E. L. (1999). Statistical learning of tone sequences by human infants and adults. *Cognition*, 70, 27-52.

Saffran, J. R. & Wilson, D. P. (2003). From syllables to syntax: multilevel statistical learning by 12-month-old infants. *Infancy*, 4, 273-284.

Schacter, D.L. (1997). The cognitive neuroscience of memory: perspectives from neuroimaging research. *Philos. Trans. R. Soc. London Ser. B*, 352, 1689–95.

Schmaltz, L.W., & Isaacson, R.L. (1968). Effects of caudate and frontal lesions on retention and relearning of a DRL schedule. *Journal of Comparative and Physiological Psychology*, 65(2), 343.

Schneider, B.A., Ghose G.M. (2012). Temporal production signals in parietal cortex. *PLoS Biology* 10(10), e1001413.

Schultz, W. (1998). Predictive reward signal of dopamine neurons. *J. Neurophysiol.*, 80, 1–27.

Schultz, W., Apicella, P., Ljungberg, T. (1993). Responses of monkey dopamine neurons to reward and conditioned stimuli during successive steps of learning a delayed response task. *J. Neurosci.* 13, 900–13.

Schultz, W., Dickinson, A. (2000). Neuronal coding of prediction errors. *Annu. Rev. Neurosci.*, 23, 473–500.

Schultz, W., Dayan, P., Montague, P.R. (1997). A neural substrate of prediction and reward. *Science* 275, 1593–99.

Shallice T. (1982). Specific impairments of planning. Philos. *Trans. R. Soc. London Ser. B*, 298, 199–209.

Shallice, T. & Burgess, P. (1991). Higher-order cognitive impairments and frontal lobe lesions in man. In H.S. Levin, H.M. Eisenberg & A.L. Benton (Eds.), *Frontal lobe function and dysfunction* (Ch. 6, pp.125-138). New York: Oxford University Press.

Shallice, T. & Burgess, P. (1996). The domain of supervisory processes and temporal organization of behaviour. *Philos. Trans. R. Soc. London Ser. B,* 351, 1405–11.

Sherman, S.M., & Guillery, R.W. (2001). Exploring the thalamus. San Diego: Academic Press.

Serre, T., Kouh, M., Cadieu, C., Knoblich, U., Kreiman, G., & Poggio, T. (2005). A Theory of Object Recognition: Computations and Circuits in the Feedforward Path of the Ventral Stream in Primate Visual Cortex. *MIT AI Memo 2005-036 / CBCL Memo 259, AI Memo 2005-036 / CBCL Memo 259 2005*. Cambridge, MA.

Serre, T., Wolf, L., Bileschi, S., Riesenhuber, M., & Poggio, T. (2007). Object recognition with cortex-like mechanisms. *IEEE Transactions on Pattern Analysis and Machine Intelligence* 29(3), 411–426.

Serre, T., Oliva, A., & Poggio, T. (2007). A feedforward architecture accounts for rapid categorization. *Proceedings of the National Academy of Sciences* 104(15), 6424–6429.

Stewart, T.C., Bekolay, T., & Eliasmith, C. (2012). Learning to select actions with spiking neurons in the basal ganglia. *Frontiers in neuroscience*, 6.

Strick, P.L. (1976). Anatomical analysis of ventrolateral thalamic input to primate motor cortex. *Journal of neurophysiology*, 39(5), 1020-1031.

Srinivasan, M. V., Laughlin, S. B. & Dubs A. (1982). Predictive coding: A fresh view of inhibition in the retina. *Proc. R. Soc. Lond. B Biol. Sci.*, 216, 427–459.

Sutton, R.S. (1988). Learning to predict by the method of temporal differences. *Machine Learning*, 3, 9–44.

Sutton, R.S., & Barto, A.G. (1998). *Reinforcement learning: An introduction*. Cambridge, MA: MIT Press.

Swanson, L.W. (2000). Cerebral hemisphere regulation of motivated behavior. *Brain research*, 886(1), 113-164.

Tanaka, K. (1996). Inferotemporal cortex and object vision. *Annual Review of Neuroscience*, 19, 109-139.

Thornton, J., Main, L. & Srbic, A. (2012). Fixed Frame Temporal Pooling. *Lecture Notes in Computer Science*. 7691, 707-718.

Todorovic, A., van Ede, F., Maris, E., & de Lange, F.P. (2011). Prior expectation mediates neural adaptation to repeated sounds in the auditory cortex: an MEG study. *Journal of Neuroscience*, 31, 9118–9123.

Tsunoda, K., Yamane, Y., Nishizaki, M. & Tanifuji, M. (2001). Complex objects are represented in macaque inferotemporal cortex by the combination of feature columns. *Nature Neuroscience*, 4, 832–838.

Tyler, R.H. (1969). Disorders of visual scanning with frontal lobe lesions. In: *Modern Neurology*, edited by Locke, S., pp.381-393. Boston: Little, Brown & Co.

Urakubo, H., Honda, M., Froemke, R.C., & Kuroda, S. (2008). Requirement of an allosteric kinetics of NMDA receptors for spike timing-dependent plasticity. *The Journal of Neuroscience*, 28(13), 3310–3323.

Vendrell, P., Junque, C., Pujol, J., Jurado, M.A., Molet, J., Grafman, J. (1995). The role of prefrontal regions in the Stroop task. *Neuropsychologia*, 33, 341–52.

Venkatraman, V., Rosati, A. G., Taren, A. A., & Huettel, S. A. (2009). Resolving response, decision, and strategic control: evidence for a functional topography in dorsomedial prefrontal cortex. *The Journal of Neuroscience*, 29(42), 13158-13164.

Wacongne, C., Labyt, E., van Wassenhove, V., Bekinschtein, T., Naccache, L., & Dehaene, S. (2011). Evidence for a hierarchy of predictions and prediction errors in human cortex. *Proc. Natl. Acad. Sci. USA* 108, 20754–20759.

Wallis, G. & Bulthoff, H.H. (2001). Effects of temporal association on recognition memory. *Proceedings of the National Academy of Science*, 98, 4800–4804.

Wallis, J.D., Anderson, K.C., & Miller, E.K. (2001). Single neurons in prefrontal cortex encode abstract rules. *Nature*, 411(6840), 953-956.

Wallis, G. & Rolls, E. (1997). Invariant face and object recognition in the visual system. *Progress in Neurobiology*, 51, 167-194.

Wang, X.J. (1999). Synaptic basis of cortical persistent activity: The importance of NMDA receptors to working memory. *Journal of Neuroscience*, 19, 9587.

Watanabe, M. (1996). Reward expectancy in primate prefrontal neurons. *Nature*, 382, 629–32.

Wikmark, R.G.E., Divac, I., & Weiss, R. (1973). Retention of spatial delayed alternation in rats with lesions in the frontal lobes. *Brain, behavior and evolution*, 8(5), 329-339.

Willshaw, D. & Dayan, P. (1990). Optimal plasticity from matrix memories: what goes up must come down. *Neural Computation* 2, 85-93.

Wood, J.N., & Grafman, J. (2003). Human prefrontal cortex: processing and representational perspectives. *Nature Reviews Neuroscience*, 4(2), 139-147.

Wurtz, R.H., Mohler, C.W. (1976). Enhancement of visual responses in monkey striate cortex and frontal eye fields. *J. Neurophysiol.*, 39, 766–72.

Yener, G. G., & Zaffos, A. (1999). Memory and the frontal lobes. *The human frontal lobes: Functions and disorders*, 288-303.

Zhang, Y., Meyers, E., Bichot, N., Serre, T., Poggio, T., & Desimone, R. (2011). Object decoding with attention in inferior temporal cortex. *Proceedings of the National Academy of Sciences*, 108(21), 8850-8855.

Zhou, X. & Luo, Y. (2013). *Implementation of Hierarchical Temporal Memory on a Many-core Architecture.* Ph.D. dissertation, Halmstad University.

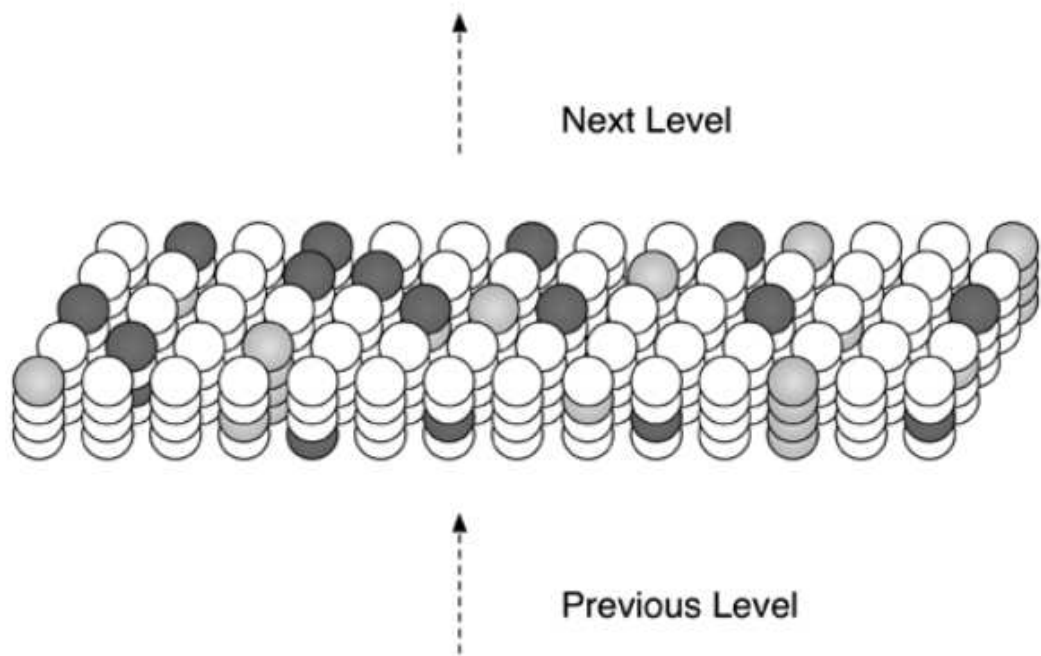

*Figure 1*. CLA columns. Each cell in a particular column represents the same feedforward input, but within a different temporal context. Active cells are shown in light grey, predictive cells in dark grey. If a column is activated by feedforward input when it had one or more cells in predictive state, then only those cells become active. Otherwise, all cells in the column are activated. Adapted from Hawkins et al, 2010.

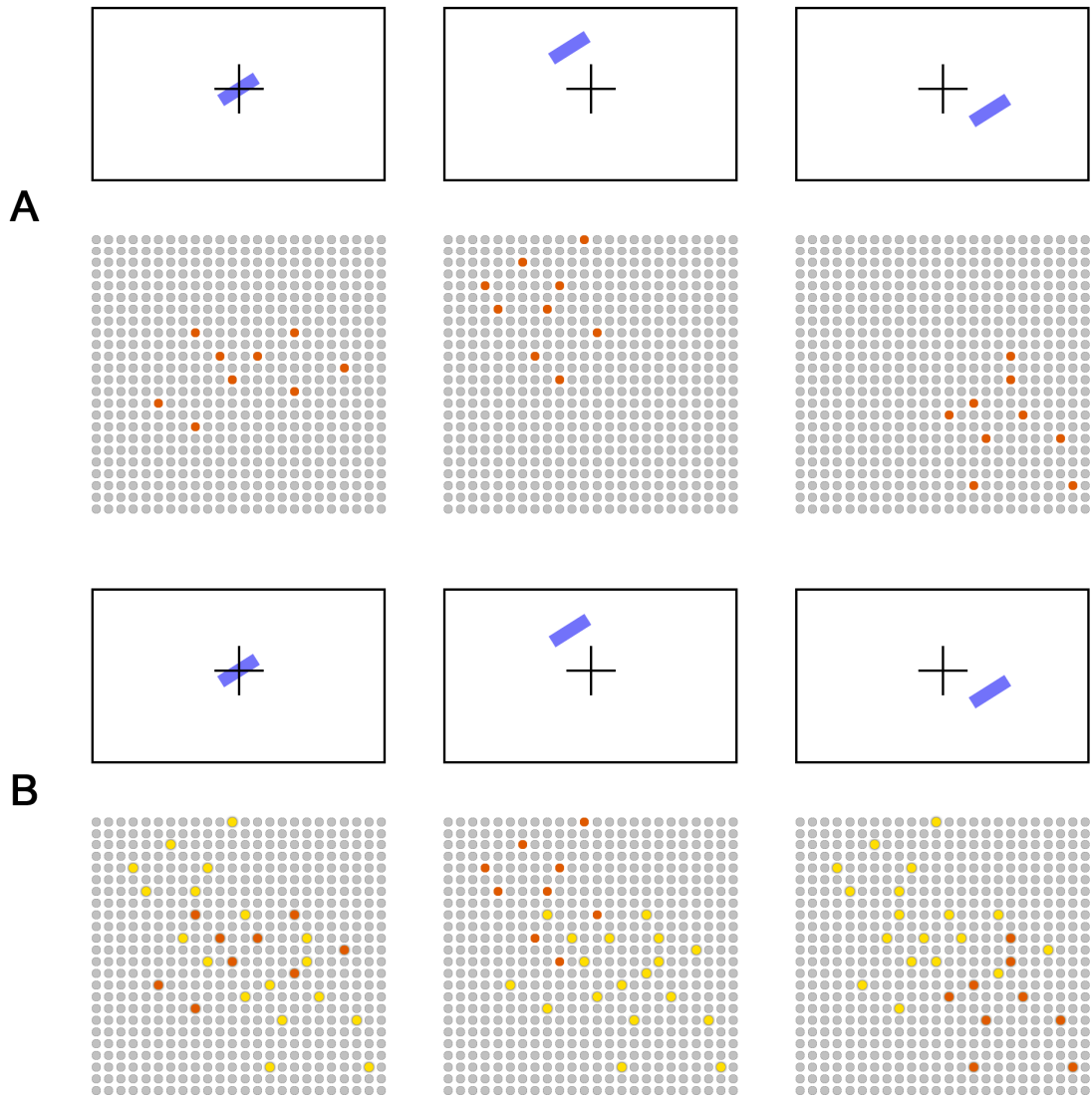

*Figure 2.* CLA visual invariance. A) The same feature appearing at different locations in the visual field results in different sparse distributed patterns of activity (orange) due to competitive Hebbian learning. Because the different spatial locations of this same feature would tend to occur sequentially with high probability, the CLA would learn to predict each of them when it receives any of them as input. B) After learning, when the CLA receives one of the spatial locations of this feature as input, it will predictively activate (yellow) representations of the others as well. This results in a degree of overlap between the representations of this feature at different nearby spatial location.

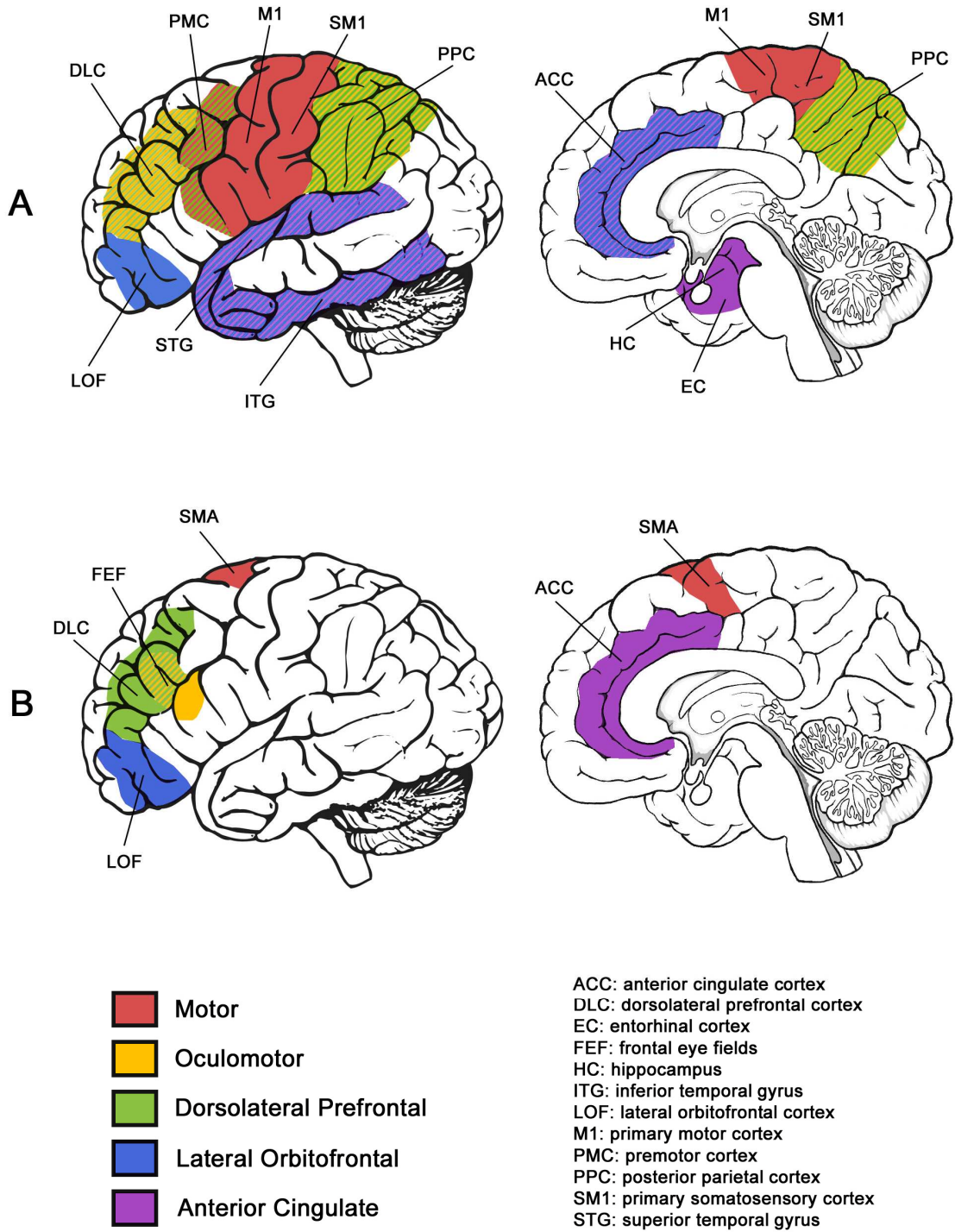

ACC: anterior cingulate cortex
DLC: dorsolateral prefrontal cortex
EC: entorhinal cortex
FEF: frontal eye fields
HC: hippocampus
ITG: inferior temporal gyrus
LOF: lateral orbitofrontal cortex
M1: primary motor cortex
PMC: premotor cortex
PPC: posterior parietal cortex
SM1: primary somatosensory cortex
STG: superior temporal gyrus

*Figure 3.* Parallel frontal cortico-thalamic relays. A) Cortical areas that project to the striatum, providing information that is used to control each relay. B) The target area of each cortico-thalamic relay. Note that information from across the cortex is used to control the relays, but all of the relay targets are frontal areas.

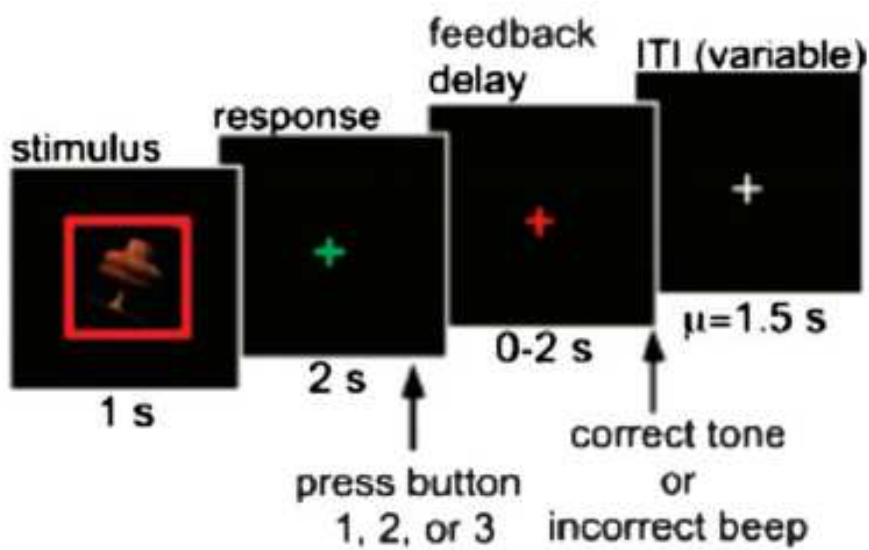

*Figure 4.* Sequence of events in the Badre et al (2010) hierarchical reinforcement learning task. Presentation of stimulus is followed by a green fixation cross. The stimulus is composed of one of three object shapes at one of three orientations, framed in a square of one of two colors. After the participant responds by pressing one of three buttons, there is a variable delay followed by auditory feedback indicating whether the response was correct. Adapted from Frank & Badre, 2012.

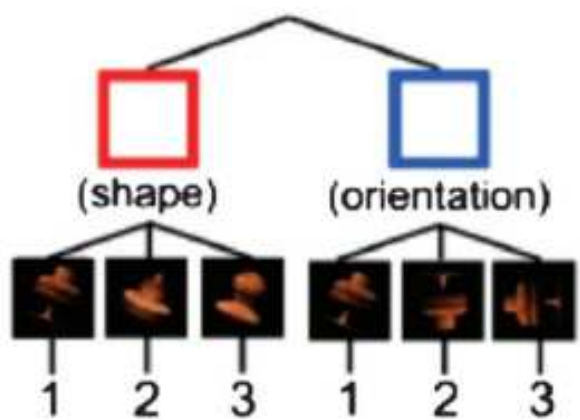

*Figure 5.* Badre et al (2010) hierarchical task condition. While stimulus-response mappings are arbitrary in the flat condition, in the hierarchical condition the mappings are dependant on the color of the framing square. In this example, in the presence of a red square, only object shape determines the mapping; in the presence of a blue square, only object orientation determines the mapping. Adapted from Frank & Badre, 2012.

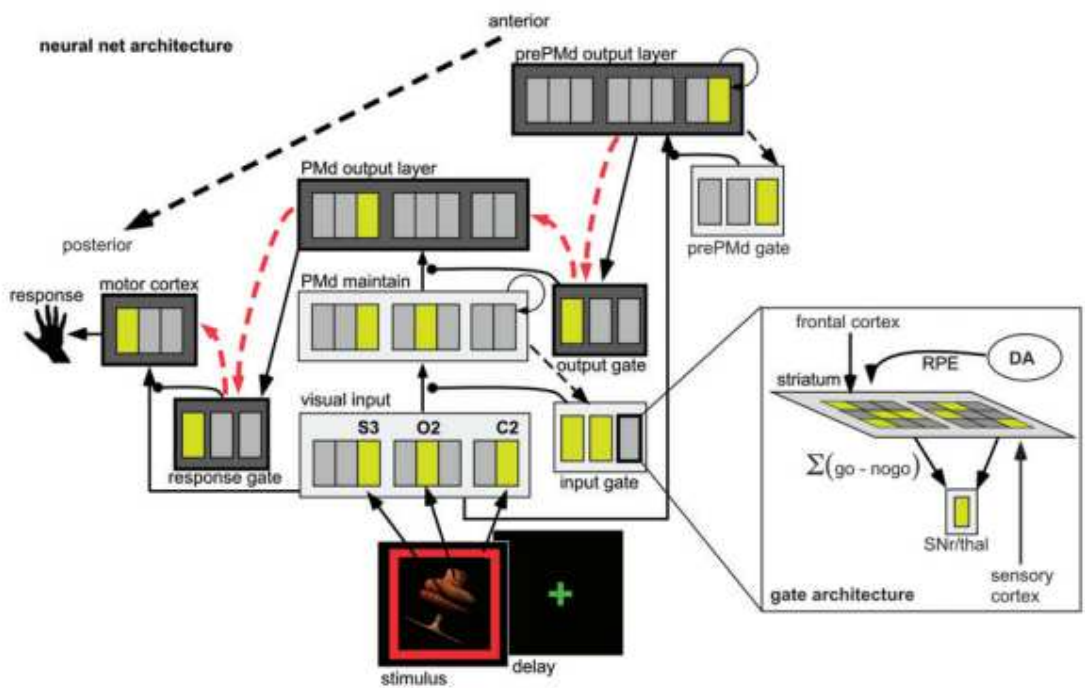

*Figure 6.* An example state of the Frank & Badre (2012) network. Arrows correspond to projections, circles correspond to basal ganglia gating effects, and dashed red arrows reflect hierarchical flow of control. Stimulus shape, orientation, and square color is represented in the visual input area. The input gate has learned to gate shape and orientation information into PMd maintain, and the prePMd gate has learned to gate square color into prePMd output. The output gate has learned that when color 2 is maintained in prePMd output, the shape information should be gated through from PMd maintain to PMd output. The response gate has learned to select button press 1 in response to the context of shape 3, which it receives from PMd output. Adapted from Frank & Badre, 2012.